\documentclass{article}
\usepackage{graphicx} 
\usepackage{booktabs} 
\usepackage{amsmath}             
\usepackage{graphicx}            
\usepackage{amssymb}
\usepackage{appendix}
\usepackage{slashed}
\usepackage{empheq}
\usepackage{comment}
\usepackage{enumerate,mdwlist}
\usepackage{scalerel,stackengine}
\def\apeqA{\SavedStyle\sim}
\def\apeq{\setstackgap{L}{\dimexpr.5pt+1.5\LMpt}\ensurestackMath{%
  \ThisStyle{\mathrel{\Centerstack{{\apeqA} {\apeqA} {\apeqA}}}}}}

\usepackage{graphicx}
\usepackage{verbatim}
\usepackage{amssymb}
\usepackage{amsmath}
\usepackage{amsfonts}
\usepackage{empheq}
\usepackage{hyperref}

\usepackage[x11names, rgb]{xcolor}
\usepackage[utf8]{inputenc}
\usepackage{mathtools}

\usepackage{geometry}

\begin{document}

\title{The Geometry of Reduction: Model Embedding, Compound Reduction, and Overlapping State Space Domains}
\author{Joshua Rosaler}
\date{}

\maketitle

\begin{abstract}
The relationship according to which one physical theory encompasses the domain of empirical validity of another is widely known as ``reduction." Here it is argued that one popular methodology for showing that one theory reduces to another, associated with the so-called ``Bronstein cube" of physical theories, rests on an over-simplified characterization of the type of mathematical relationship between theories that typically underpins reduction. An alternative methodology, based on a certain simple geometrical relationship between distinct state space models of the same physical system, is then described and illustrated with examples. Within this approach, it is shown how and under what conditions inter-model reductions involving distinct model pairs can be composed or chained together to yield a direct reduction between theoretically remote descriptions of the same system. Building on this analysis, we consider cases in which a single reduction between two models may be effected via distinct composite reductions differing in their intermediate layer of description, and motivate a set of formal consistency requirements on the mappings between model state spaces and on the subsets of the model state spaces that characterize such reductions. These constraints are explicitly shown to hold in the reduction of a non-relativistic classical model to a model of relativistic quantum mechanics, which may be effected via distinct composite reductions in which the intermediate layer of description is either a model of non-relativistic quantum mechanics or of relativistic classical mechanics. Some brief speculations are offered as to whether and how this sort of consistency requirement between distinct composite reductions might serve to constrain the relationship that any unification of the Standard Model with general relativity must bear to these theories.
\end{abstract}

\section{Introduction}


Common lore dictates that theories in physics progress toward increasing generality and unification. Galileo's theory of terrestrial gravitation gave way to Newton's more encompassing Universal Theory of Gravitation, which in turn gave way to Einstein's still more encompassing theory of general relativity, which, it is expected, will in turn give way to some even more encompassing quantum theory of gravitation. Similarly, the theory of electrostatics gave way to Maxwellian electrodynamics, which gave way to quantum electrodynamics, which gave way to the Standard Model, which in turn, it is expected, will give way to a theory that also encompasses gravitational phenomena. The relationship between any pair of successive theories in such a sequence, whereby the later theory encompasses the domain of empirical validity of the earlier theory, is widely known as ``reduction." Reduction is the link between successive theories on which claims of increasing unification in physics are ultimately rooted. Indeed, reduction of general relativity is part of what it means to \textit{be} a theory of quantum gravity. 
\footnote{For extended discussion of this point, see Crowther \cite{crowther2017inter}.
}
Likewise, reduction of the Standard Model is a necessary requirement on beyond the Standard Model (BSM) theories, and the need to simultaneously reduce both general relativity and the Standard Model is a definitional requirement on any ``theory of everything." 

In many cases, the conventional wisdom of steadily increasing unification in physics implicitly takes for granted reduction between certain established theories without explicit proof, ostensibly on the basis of a certain philosophically motivated faith in the unity of nature. Nevertheless, the detailed mathematical relationships between theories that underpin reduction in many cases are far from trivial, as in the case of the reduction between classical and quantum theories, where concerns about quantum measurement, decoherence, the $\hbar \rightarrow 0$ limit, Ehrenfest's Theorem, and other results all appear to play some role. Likewise, concerns about renormalization make the reduction of quantum mechanics to quantum field theory,
\footnote{There exist conflicting conventions as to the precise use of the term ``reduction." On one usage, it is the more encompassing theory that ``reduces to" the less encompassing theory; such uses reflect the understanding of reduction as simplification, as in ``6/4 reduces to 3/2." On another set of usages, it is rather the less encompassing theory that ``reduces to" the less encompassing theory; such uses reflect the understanding of reduction as subsumption into a more general framework, as in the claim that ``chemistry reduces to physics" or ``thermodynamics reduces to statistical mechanics." We will adopt the latter convention here, so that the less encompassing description is understood to reduce to the more encompassing description. For further discussion of the distinction between these uses of the term ``reduce," see Nickles' \cite{nickles1973two}.
}
 which is also widely taken for granted, more subtle than it may at first appear. It is reasonable to expect that lessons gleaned from more careful study of relationships between established theories may be useful in the study of relationships between speculative models of new physics and current theories. For example, it is reasonable to expect that a clearer understanding of decoherence, renormalization, and quantum measurement, and of the manner in which these elements work together will have some role to play in understanding the connection between already established theories and whatever the correct theory of quantum gravity turns out to be. 

The present discussion seeks to further develop the methods of reduction elaborated in \cite{Rosaler2018}, \cite{rosaler2015local}, and \cite{RosalerThesis}. Section \ref{Bronstein} rehearses the reasons why one popular approach to problems of reduction, associated with the ``Bronstein cube" of physical theories and based on the notion that reduction consists simply in taking limits of constants of nature, rests on a picture of the mathematical requirements for reduction that is either excessively vague or  incorrect. Section \ref{BridgeFunctions} reviews the core elements of one alternative approach, based on the notion that reduction fundamentally concerns relationships between distinct models of the same physical system, and requires the identification of quantities within the reducing model that approximately instantiate the physically salient structures of the reduced model over a restricted subset of the reducing model's state space. Where relations between the models of fundamental physics 
\footnote{Theories of ``fundamental physics" here are understood to include non-statistical mechanical and non-thermodynamic theories, such as classical mechanics, quantum mechanics, relativistic quantum mechanics, quantum field theory, and quantum gravity. While many of the claims about reduction defended here also apply in the context of thermodynamics and statistical mechanics, these theories introduce novel statistical aspects that demand special treatment. 
}
are concerned, this instantiation can be shown to hold by virtue of a certain geometrical relationship between dynamical group actions over the state spaces of the two models. Section \ref{ReductionCommute} advances the central claims of the article, which concern the manner in which distinct inter-model reductions may be composed, and advances a set of consistency requirements in cases where a single reduction may be effected via multiple distinct compound reductions differing in their intermediate layer of description. This consistency requirement is checked in the special case of the reduction of non-relativistic classical mechanics to relativistic quantum mechanics, which may be effected via an intermediate model either of non-relativistic quantum mechanics or of classical relativistic mechanics. On the view that reduction is simply of matter of taking some limit, it is expected that the classical limit $\hbar \rightarrow 0$ limit should commute with the non-relativistic limit $c \rightarrow \infty$, although presentations of the Bronstein cube typically do not specify whether or in what sense this might actually turn out to be the case, or which specific quantities one should be taking the limit of; the consistency condition described here serves to illustrate more explicitly one important sense in which the quantum-to-classical and relativistic-to-non-relativistic transitions commute with each other. Some brief speculative remarks are then offered as to how these consistency conditions might serve to constrain the relationship between potential models of new physics and models drawn from established theories. Section \ref{Conclusion} is the Conclusion.

\begin{figure}
   \includegraphics[width=0.35\textwidth]{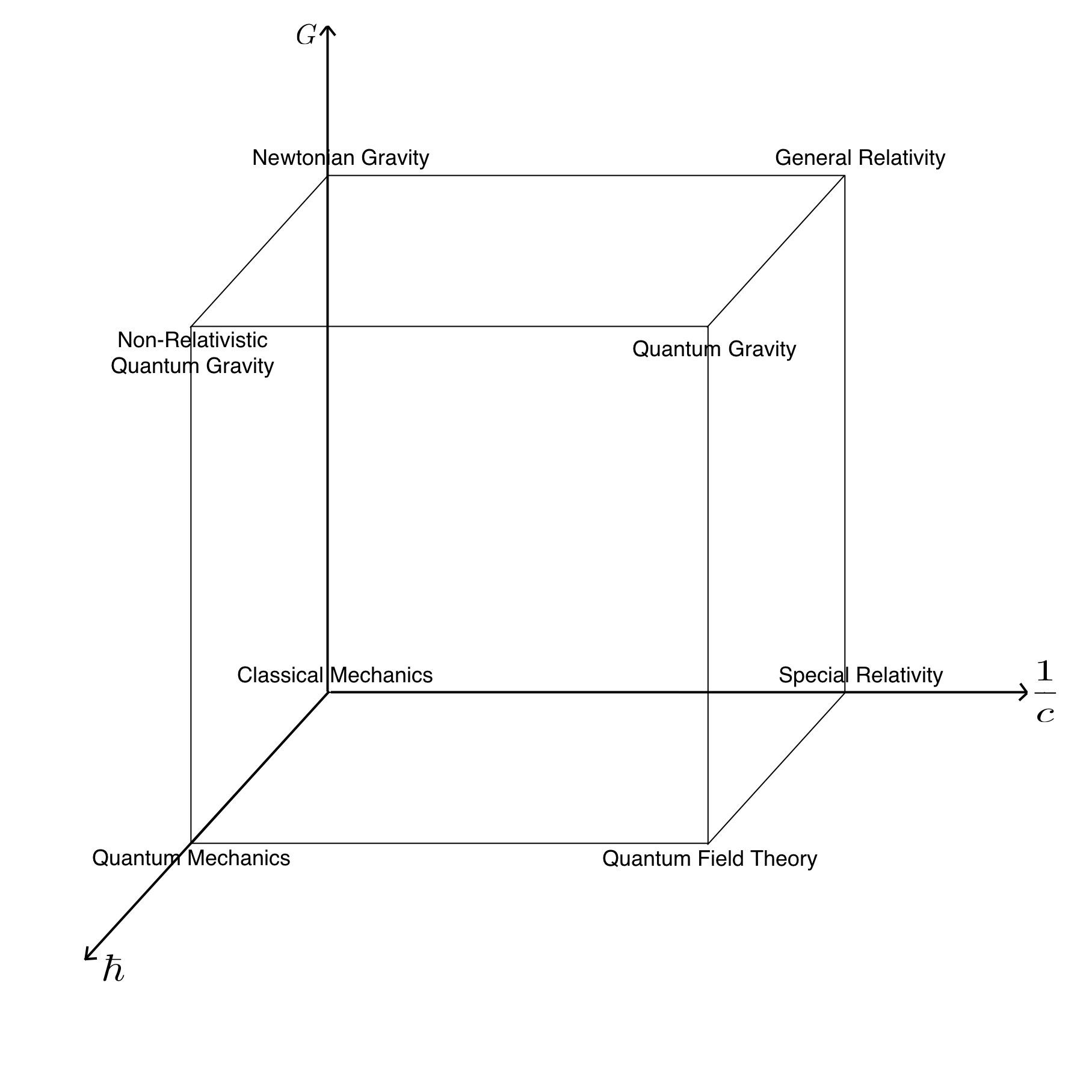}
   \hfill
   \includegraphics[width=0.6\textwidth]{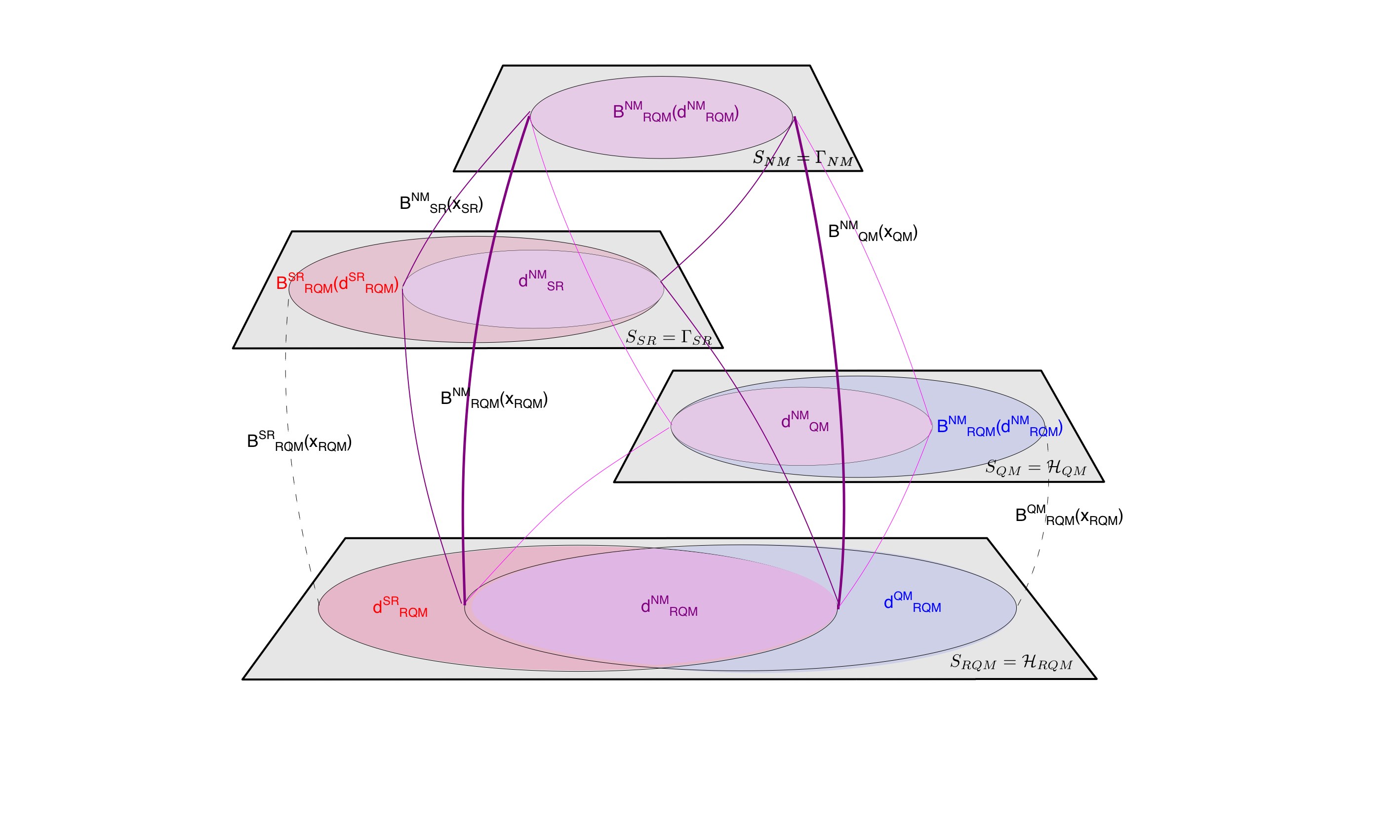}
   \caption{\footnotesize Two distinct strategies for mapping the links between theories of fundamental physics, compared in the discussion below. On the left is the so-called Bronstein cube of physical theories, described in Section \ref{Bronstein}. On the right is an alternative picture based on a certain geometrical relationship between state space models, described in Sections \ref{BridgeFunctions} and \ref{ReductionCommute}. The image on the right, which concerns the relationship between the classical and non-relativistic domains of relativistic quantum theory, corresponds roughly to the bottom face of the Bronstein cube.
 }
\end{figure}

\section{The Role of Limits in Reduction} \label{Bronstein}

Limiting relations, in which the mathematical structures of the reduced theory are recovered as some parameter in the reducing theory is varied, play a crucial role in many reductions: the laws of Newtonian mechanics are recovered as an approximation to those of special relativity for speeds much less than that of light; effective field theories describing some species of light particle are recovered as an approximation to more encompassing effective field theories in the limit where other particle species are much more massive; thermodynamic regularities are recovered as an approximation to those of statistical mechanics when the number of degrees of freedom becomes very large; the list continues. The prevalence of limits across so many cases has given rise to a manner of speaking in which reduction is designated simply as the requirement that one theory be a limit or limiting case of another. That is, reduction is sometimes taken to require, seemingly \textit{as a matter of definition}, that one theory be a mathematical limit of another. Yet, given that the ultimate aim of reduction is to show that all physical systems that can be modeled in the reduced theory can be modeled more precisely in the reducing theory (which also should incorporate phenomena outside the scope of the reduced theory) it is important to keep in mind that limiting relations are merely a means to this end, rather than the ultimate goal of reduction. For this reason, it has recently been argued that one approach to reduction, based on the notion that reduction is essentially about showing one theory to be a limit of another, and exemplified by the so-called ``Bronstein cube" of physical theories, rests on a flawed understanding of the mathematical relationship between theories according to which one theory encompasses the domain of empirical validity of another. 

\begin{figure} \label{BronsteinFig}
\center{ \includegraphics[width=0.4\textwidth]{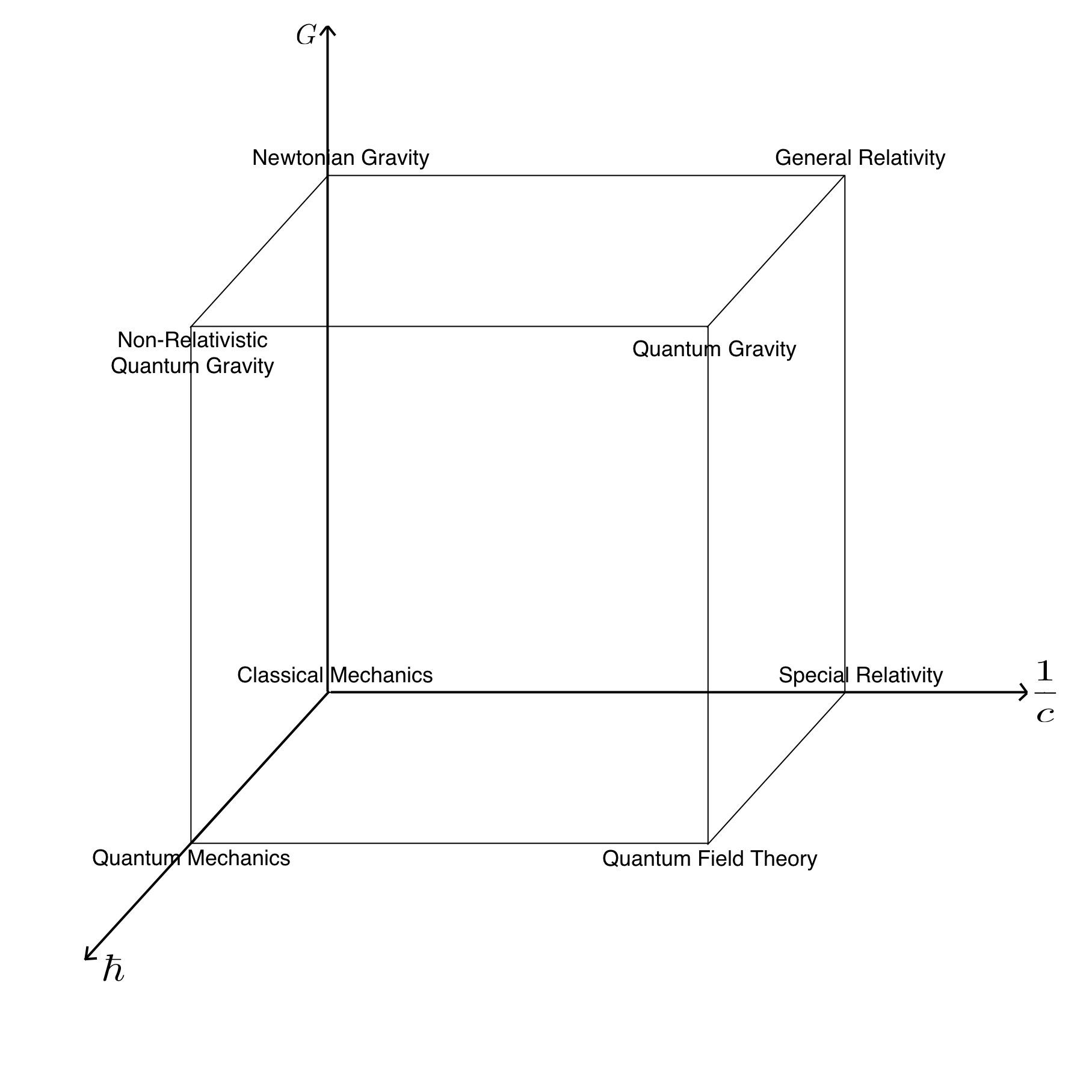} }
\caption{The Bronstein cube of physical theories, according to which passage between physical theories is effected by limits as different constants of nature, associated with distinct axes, are varied. }
\end{figure}

As shown in Figure \ref{BronsteinFig}, the Bronstein cube places the theories of modern physics - including some as yet undiscovered theory of quantum gravity, which is understood to encompass the empirical domains of validity of both general relativity and quantum field theory - at the corners of a cube, where passage between theories at opposite ends of a given  edge is effected through a mathematical limit in which some constant of nature is varied. According to the figure, transition from quantum to classical theories, represented by passage along the four edges parallel to the $\hbar$ axis, is effected by taking the limit $\hbar \rightarrow$ of vanishing Planck's constant. Transition from relativistic to non-relativistic theories, represented by passage along the four horizontal edges, is effected by taking the limit $c \rightarrow \infty$ in which the speed of light becomes infinite. Transition from gravitational to non-gravitational theories, represented by passage along the four vertical edges, is effected by taking the limit $G \rightarrow 0$ of vanishing Newton's constant. The Bronstein cube is thought to originate in a paper by Gamow,  Ivanenko, and Landau, \cite{gamow2002world}, and is further discussed in \cite{stachel2005development}, \cite{duff2002trialogue}, \cite{christian1997exactly}. Recently, Oriti has proposed an extension of the cube, which he calls the ``Bronstein hypercube," to include a fourth axis corresponding to the number $N$ of degrees of freedom in a system \cite{oriti2018bronstein}.

The difficulties with this way of thinking about the relationships among the theories of modern physics have been discussed extensively elsewhere, so we review them only briefly here. Oriti himself, a leading authority on the Bronstein cube, writes that it provides only ``an extremely rough sketch of theoretical physics." As he explains, ``It does not account even remotely for the complexity of phenomena that are actually described by the mentioned frameworks. And it does not say anything about the very many subtleties involved in actually moving from one framework to the other, and back from there" \cite{oriti2018bronstein}. Moreover, the picture of inter-theory reduction provided by the cube is extremely vague in its requirements. Apart from the demand that some limit be involved in the transition between theoretical frameworks, this approach does not offer any clear recipe for determining whether reduction holds in a given case. Given a pair of theories, how precisely does one determine whether one is a limit of the other, or whether one reduces to the other? Does the existence of any limiting relationship between any two quantities of the theories suffice to establish one theory as a limit of another? Or is it necessary that these limits involve specific types of quantities within the theories? If the latter, specifically what sorts of quantities in the reduced theory must be recovered as limits of quantities in the reduced theory? Does every quantity in the reducing theory need to go over in the limit to some quantity in the reduced theory? If we interpret the limits of the cube naively and literally, it is clear that the limiting relations suggested by the cube do not hold; for example, simply taking the $\hbar \rightarrow 0$ limit of Schrodinger's equation gives nonsense, not classical mechanics. More needs to be said about how one decides, given a pair of theories, whether one is a limit of another, and also how this serves to ensure subsumption of one theory's domain by another. If the sole contents of the Bronstein cube is that the limits associated with edges of the cube are somehow related to the task of showing that one theory encompass the domain of empirical validity of another, then it must be regarded as more of a vague heuristic than a well-formulated account of the requirements for reduction. In attempting to understand how one theory encompasses the domain of empirical validity of another, much more needs to be said in the context of concrete systems about how the reducing theory serves to represent those physical degrees of freedom that are well-described by the reduced theory.


A second worry, apart from vagueness, concerns whether the limits in the diagram commute, as the diagram indicates that they should \cite{HossiBronstein} and \cite{ButterfieldQGTalk}. For example, the diagram indicates that it should be possible to pass, say, from (relativistic) quantum field theory to Newtonian mechanics, by first taking the limit $\hbar \rightarrow 0$ and then the limit $c \rightarrow \infty$, or alternatively by first taking the limit $c \rightarrow \infty$ and then the limit $\hbar \rightarrow 0$. On the other hand, it means little to say that these limits should commute until one specifies the particular set of $\hbar$- and $c$-dependent quantities that one is taking these limits \textit{of}. Many presentations of the Bronstein cube approach neglect to specify what these quantities are or how they should be determined for a given pair of theories.  In the next sections, we will see how the notion that different reductions commute can be made more precise. 

A third concern is that the limits are based on varying constants of nature, which are fixed for real systems; thus, it is sometimes argued that the relevance of such limiting relations for real systems is obscure \cite{HossiBronstein} and \cite{ButterfieldQGTalk}. However, a recent effort to address this second objection in the context of the limit $\hbar \rightarrow 0$ has been made by Feintzeig, and in the context of the limit $c \rightarrow \infty$ by Fletcher, who attempt to interpret these limits not as counterfactual changes in the constants of nature, but as changes of the units in which the numerical values of these constants are given \cite{feintzeig2018status}. 

A fourth objection, expounded at length in \cite{rosaler2017reduction}, is that the approach to reduction based on the Bronstein cube treats reduction as a purely formal mathematical relationship, in the sense that it should be possible to determine whether one theory reduces to another given knowledge only of their mathematical frameworks. This view arises naturally from the tendency to see limiting relations as the end goal of reduction. However, recalling that reduction requires one theory to subsume the domain of empirical validity of another, it is clear that knowledge of the mathematical frameworks of the theories alone is \textit{not} sufficient to determine whether one reduces to the other. It is also necessary to have some empirical knowledge of the domain of empirical success of the reduced theory, and of the precision with which it describes systems in its domain, so that one knows just how closely and in what specific contexts the mathematical frameworks of the theories need to dovetail in order to ensure that one subsumes the domain of the other. 

The Bronstein cube, and the notion that reduction is simply a matter of taking limits of constants of nature, is useful perhaps as a rough, partial heuristic. Here I wish simply to warn against the temptation to take the picture of reduction suggested by the cube too seriously, and to conclude merely by virtue of the existence of some formal limiting relation between theories that reduction has been shown to hold. Particularly in the context of quantum-classical relations, it seems clear that the limit $\hbar \rightarrow 0$ alone does not suffice for reduction since complex mechanisms associated with decoherence and quantum measurement also must play a role in reduction in these cases.

\section{Reduction between State Space Models} \label{BridgeFunctions}


The methodology described below can be understood as an attempt formalize the characterization of reduction contained in the following quotation from Wallace, in which reduction is the requirement that the more fundamental, reducing theory \textit{instantiate} the physically salient structures of the reduced theory in contexts where the latter is successful:

\begin{quote}
This instantiation relation (I claim) is the right way of understanding the relationship between different scientific theories - the sense in which one theory may be said to ``reduce" to another.  Crucially: this ``reduction", on the instantiation model, is a local affair: it is not that one theory is a limiting case of another \textit{per se}, but that, \textit{in a particular situation}, the ``reducing" theory instantiates the ``reduced" one" \cite{wallace2012emergent}, Ch. 2. 
\end{quote}

\noindent Building on this characterization, reduction of theory $T_{h}$ to theory $T_{l}$ (where the subscripts $h$ and $l$ designate ``high-level" and ``low-level" respectively) requires that any circumstance under which the behavior of a real system $K$, understood as some set of physical degrees of freedom, is accurately modeled in $T_{h}$ be a circumstance under which it can be modeled more accurately and universally in $T_{l}$. 
Thus, reduction between theories $T_{h}$ and $T_{l}$ is grounded in reductions between specific models $M_{h}$ and $M_{l}$ of systems $K$ in $T_{h}$'s domain of empirical validity.  

Here, we understand a model $M$ of physical system $K$ to be specified by a state space $S$, which in the cases of interest to us here possesses the structure of a differentiable manifold (and often also the structure of a vector space), a notion of distance between states generated by a metric or norm defined over the manifold, and some further structure defined over the state space manifold that prescribes the dynamics of the model (e.g., Lagrangian, Hamiltonian, equations of motion). 

What then does it mean for one model $M_{h}$ of some set of physical degrees of freedom $K$ to reduce to another model $M_{l}$? It means that every circumstance in which the physical degrees of freedom $K$ are well modeled by $M_{h}$ is a circumstance in which these same degrees of freedom are modeled more accurately by $M_{l}$. However, the state spaces of $M_{l}$ and $M_{h}$ may represent different degrees of freedom (dof's), where the dof's represented by the latter ``supervene on," 
\footnote{Property A (e.g., a macrostate) supervenes on property B (e.g., a microstate) if and only if there can be no difference in A without a difference in B. The value of property A uniquely determines the value of property B while the reverse is typically not the case. 
}
or are uniquely determined by, those represented by the former. This dependence is typically represented by a function $B: S_{l} \rightarrow S_{h}$ from the low-level state space $S_{l}$ to the high-level state space $S_{h}$ that establishes a mathematical bridge between the models. The quantity $B(x_{l})$, whose behavior is determined entirely by the behavior of the low-level state $x_{l} \in S_{l}$ prescribed by the low-level model $M_{l}$, specifies $M_{l}$'s representation of the degrees of freedom that are well described by $M_{h}$. For example, the center of mass of a classical composite object (such as a ball) may be modeled either by a classical Hamiltonian model whose state space directly represents the behavior of its center of mass or by a more detailed classical model describing the behavior of the ball's microscopic constituents, where the center of mass is represented instead by a particular function $B$ of the microscopic state. However, as we will see explicitly below, in cases of reduction $B(x_{l})$ typically only mimics the behavior of the high-level state $x_{h}$ prescribed by $M_{h}$ for $x_{l}$ in some restricted subset $d \subset S_{l}$. For reduction to occur, the induced trajectory $B(x_{l}(t)$ must approximate the high-level trajectory $x_{h}(t)$ roughly within the empirically determined margin of error $\delta_{emp}$ within which the high-level $x_{h}(\tau)$ is known to track the real physical behavior of the degrees of freedom $K$. Moreover, it must do so over the empirical timescale $T_{emp}$ onverwhich the high-level trajectory $x_{h}(t)$ tracks $K$ within $\delta_{emp}$.


In more general contexts, the relevant parameter characterizing dynamical flows in the state space may not necessarily be time, but some more general variable (or set of variables)  $\tau$ that parametrizes physical solutions of the model. Generally, reduction requires that for every physically realistic 
\footnote{``Physically realistic" here indicates that the trajectory approximates the behavior of the real physical degrees of freedom described by the model to within some specified error tolerance. 
}
solution $x_{h}(\tau)$ of $M_{h}$, there exist some physically realistic solution $x_{l}(\tau)$ of $M_{l}$ such that $\left| B(x_{l}(\tau)) - x_{h}(\tau) \right|_{h} < 2\delta_{emp}$ over ranges of $\tau$ for which $x_{h}(\tau)$ approximates the behavior of $K$ within $\delta_{emp}$. 
\footnote{The difference between $x_{h}(\tau)$ and $B(x_{l}(\tau))$ is required to be less than $2\delta_{emp}$ because if the $x_{h}(\tau)$ and $B(x_{l}(\tau))$ both approximate  $K$'s behavior within error bound $\delta_{emp}$, then they may differ from each other by at most $2\delta_{emp}$.
}
In the next subsection, we discuss how these requirements can be formalized in terms of flows associated with group actions over the models' respective state spaces. 

\subsection{Formal Requirements: Matching of State Space Group Actions over a Restricted State Space Domain}

We now explain how the state-space-based approach to reduction described in general terms in the previous section may be formalized as a relationship between group actions over the state spaces of the reduced and reducing models. The mathematical style of the discussion will be informal so as to place primary emphasis on the central concepts. However, we explain in broad terms how some of the more detailed technicalities can be filled in more rigorously. 

Here, we focus on the large set of cases where both high- and low-level models can be formulated in terms of first-order, deterministic dynamical equations of motion over some state space
\footnote{Many features of the approach to reduction described here can be extended straightforwardly to reductions involving models with stochastic dynamics, with certain important modifications to accommodate the probabilistic nature of the models. For example, approximate equality of induced and high-level state space trajectories is replaced by approximate equality with high likelihood. The instantiation picture of reduction may also be extended to non-dynamical models such as the  model of temperature, pressure, and volume in an Ideal Gas, although we do not explore this here. 
}:

\begin{align*}
& \frac{d x_h^{\mu}}{d \tau} = V_h^{\mu} \big|_{x_h} \\
& \frac{d x_l^{\mu}}{d \tau} = V_l^{\mu} \big|_{x_l},
\end{align*}

\noindent where $\tau$ is a flow parameter (usually time), the high-level dynamics are generated by vector field $V_{h} \big|_{x_{h}} = V_{h}^{\mu}(x_{h}) \frac{\partial}{\partial x_{h}^{\mu}} \in T_{x_{h}}S_{h}$ and the low-level dynamics by $V_{l}\big|_{x_{l}} = V_{l}^{\mu}(x_{l}) \frac{\partial}{\partial x_{l}^{\mu}} \in T_{x_{l}}S_{l}$. The $M_{h}$-prescribed evolution of an arbitrary initial high-level state $x_{h}^{0} \in S_{h}$ is then given by $x_{h}(\tau) = \left[ e^{\tau V_{h}} \ x_{h} \right]_{x_{h} = x_{h}^{0}}$. Likewise, the $M_{l}$-prescribed evolution of an arbitrary initial low-level state is given by $x_{l}(\tau) = \left[ e^{\tau V_{l}} \ x_{l} \right]_{x_{l} = x_{l}^{0}}$.

Since the degrees of freedom described by $M_{h}$ are presumed to supervene on the degrees of freedom described by $M_{l}$, there will be some function $B: S_{l} \rightarrow S_{h}$ that serves to characterize the mathematical dependence of the degrees of freedom represented by $M_{h}$ on those represented by $M_{l}$, and thereby to identify $M_{l}$'s representation, $B(x_{l})$, of the degrees of freedom described by $M_{h}$.  Informally, reduction requires that whenever $x_{h}(\tau)$ accurately describes the behavior of the system $K$, there exist a physically realistic $x_{l}(\tau)$ such that $x_{h}(\tau) \approx B(x_{l}(\tau))$, or more explicitly,

\begin{align} \label{FlowCommutation}
\left[ e^{\tau V_{h}} \ x_{h} \right]_{x_{h} = B(x_{l}^{0})}    \approx   B\left(\left[ e^{\tau V_{l}} \ x_{l} \right]_{x_{l} = x_{l}^{0}}    \right). 
\end{align}

\noindent In other words, mapping up to the high-level state space with $B$ and then applying the high-level dynamics should yield approximately the same result as applying the low-level dynamics and then applying $B$; that is, dynamical evolution should approximately commute with the function $B$. 

\begin{figure}
\center{\includegraphics[width= 0.5\linewidth]{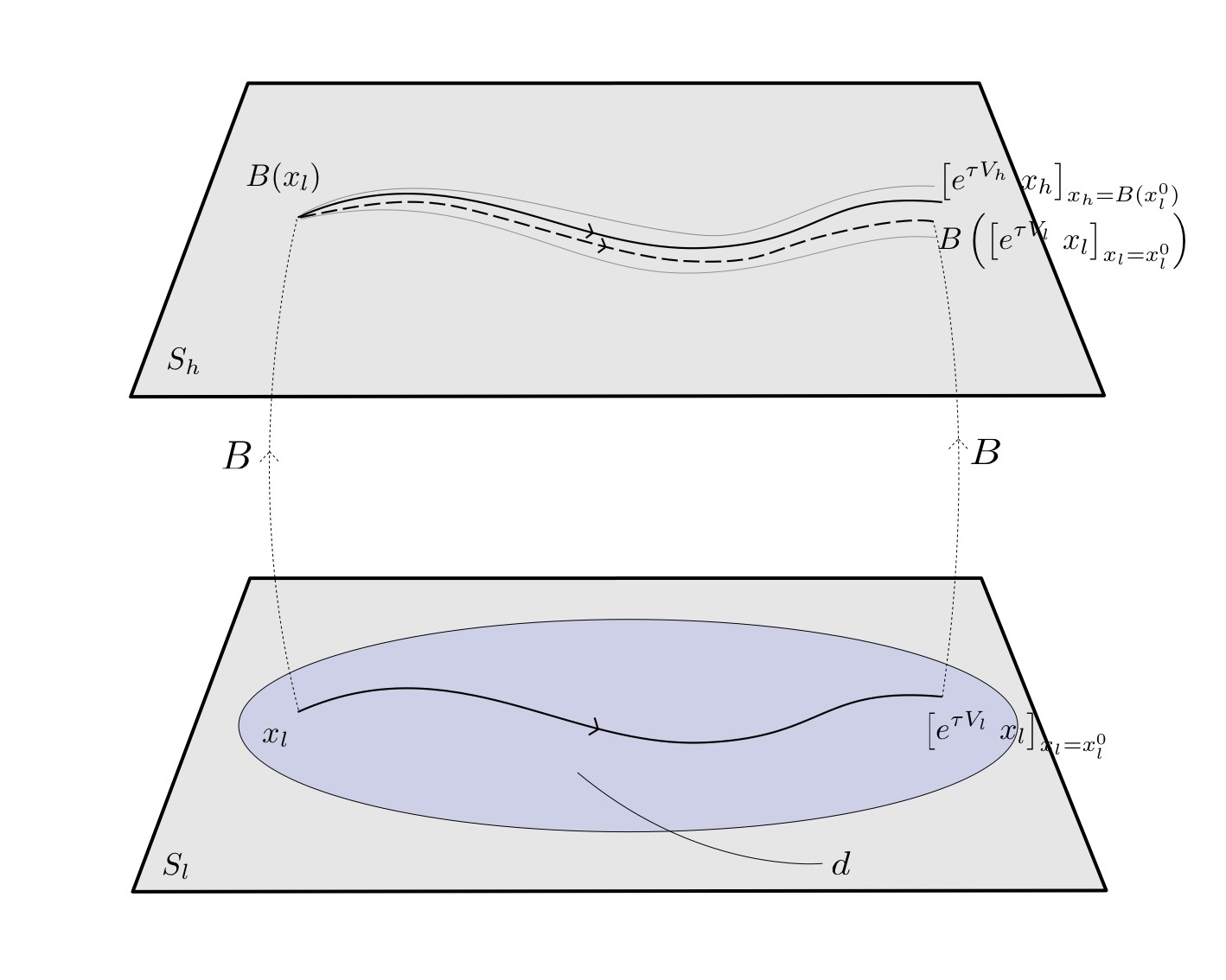}}
\caption{In cases where a single system $K$ can be described both by a high-level model $M_{h}$ and a low-level model $M_{l}$, which may describe more fundamental degrees of freedom, reduction of $M_{h}$ to $M_{l}$ requires that the quantity $B(x_{l}(\tau))$ (dotted curve in $S_{h}$) approximate $x_{h}(\tau)$ (solid curve in $S_{h}$) within empirical margin $2\delta_{emp}$ (indicated by light grey lines in $S_{h}$).  }
\label{Commute}
\end{figure}

Let us make this more precise. Consider first the notion of approximate equality implied by the symbol $\approx$ in (\ref{FlowCommutation}). When can the high-level trajectory $x_{h}(t)$ and the induced trajectory $B(x_{l}(\tau))$ be considered approximately equal? Since reduction requires that $M_{l}$'s description of the physical degrees of freedom $K$ be more accurate than $M_{h}$'s, it is sensible to require that $x_{h}(\tau)$ and $B(x_{l}(\tau))$ agree within the empirical margin of error $\delta_{emp}$ within which $M_{h}$ is known to track the behavior of $K$. Moreover, this margin should be respected over ranges $T_{emp}$ of the flow parameter $\tau$ for which the trajectory $x_{h}(\tau)$ tracks $K$ within $\delta_{emp}$. 
\footnote{For example, the deterministic equations of classical mechanics are only known to describe the trajectory of Saturn's moon Hyperion over certain limited timescales, beyond which classical predictability is lost due to quantum and classically chaotic effects. See for example \cite{zurek1995quantum} and \cite{zurek1998decoherence} for further details. Quantum models should only be required to recover classical trajectories over these timescales for which they furnish an accurate representation of the system in question. 
}
 Thus, relation (\ref{FlowCommutation}) can be stated more precisely as the requirement that

\begin{align} \label{FlowCommutationExact}
 \left|  \left[ e^{\tau V_{h}} \ x_{h} \right]_{x_{h} = B(x_{l}^{0})}   -  B\left(\left[ e^{\tau V_{l}} \ x_{l} \right]_{x_{l} = x_{l}^{0}} \right)   \right|_{h} < 2\delta_{emp} 
\end{align}

\noindent for $0 \leq \tau < T_{emp}$ (see Figure \ref{Commute}). In classical phase space, for example, the norm $\left| \ \right|_{h}$ can be taken as the Euclidean norm; in Hilbert space, it can be taken as the norm associated with the Hilbert space inner product. It is particularly important to emphasize that the relation (\ref{FlowCommutationExact}) will generally hold only for as long as the low-level state $x_{l}(\tau)$ remains in some restricted subset $d \subset S_{l}$ of the low-level state space; when it leaves this subset, the difference between $x_{h}(\tau)$ and $B(x_{l}(\tau))$ will exceed this margin. When these two representations of $K$ diverge, reduction requires the low-level model's representation $B(x_{l}(\tau))$ to be the more accurate of the two. 

It is useful to note that the condition (\ref{FlowCommutation}) will hold if the quantity $B(x_{l}(\tau))$ approximately satisfies the high-level equations of motion that are satisfied by $x_{h}(\tau)$:

\begin{equation} \label{EOMReduced}
 \boxed{
\frac{d B^{\mu}(x_{l}(\tau))}{d \tau} \approx V_{h}^{\mu}\big|_{B(x_{l}(\tau))}.
} 
\end{equation}

\begin{figure} 
\center{\includegraphics[width=.7 \linewidth]{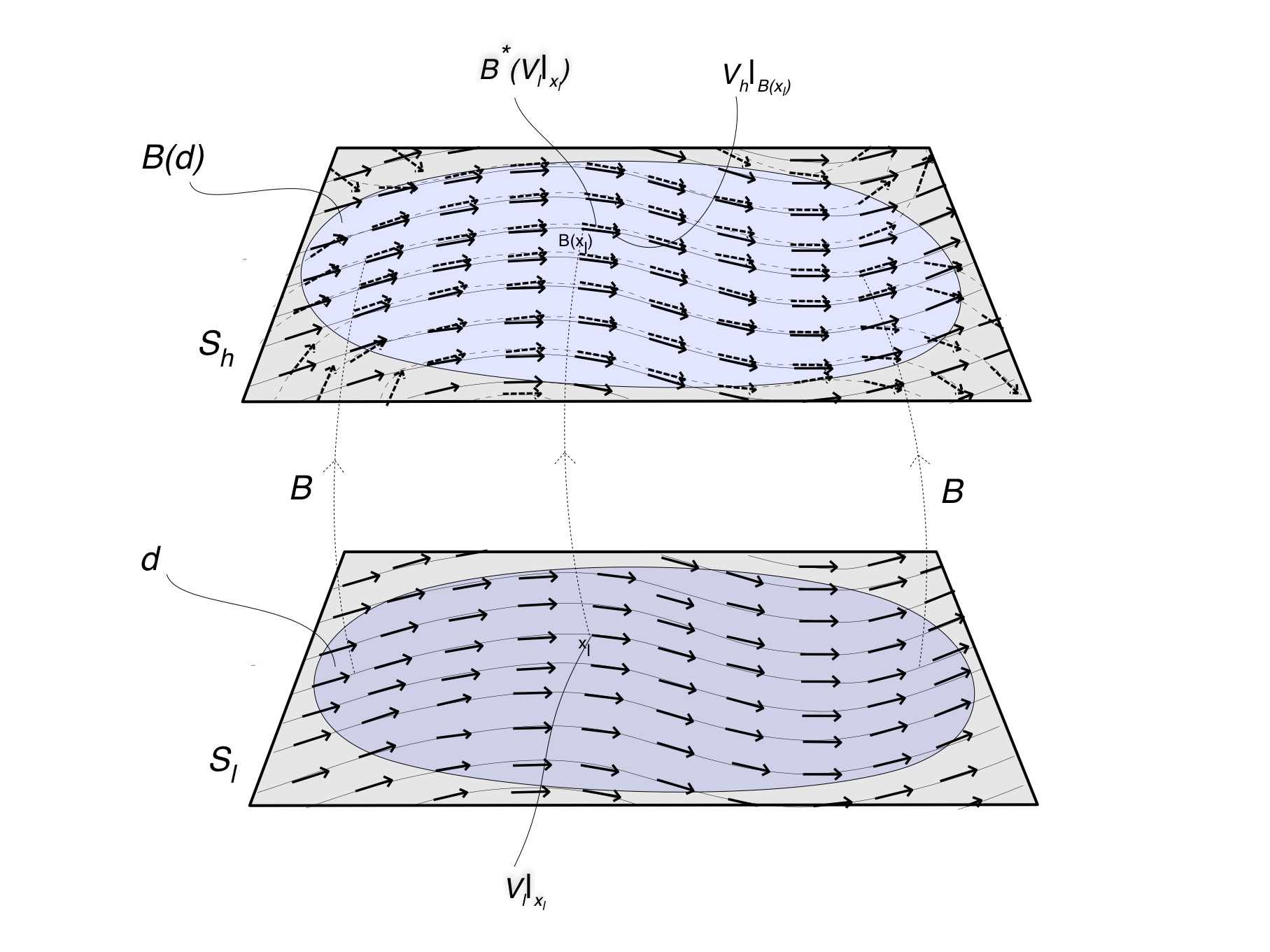}} 
\caption{If the high-level dynamical vector field $V_{h}$ evaluated at $B(x_{l})$ is approximately equal to the push-forward of the low-level vector field $V_{l}$ (dotted arrows in upper state space) evaluated at $x_{l}$, for all $x_{l} \in d$, then the trajectory induced on $S_{h}$ through $B$ by the integral curves of $V_{l}$ in $d$ will approximate the integral curves of $V_{h}$ in the image domain $B(d) \subset S_{h}$. 
}
\label{VectorFields}
\end{figure}

\noindent \footnote{One can see that (\ref{FlowCommutation}) follows from (\ref{EOMReduced}) by integrating both sides of (\ref{EOMReduced}) with respect to the parameter $\tau$. 
}
\noindent For (\ref{FlowCommutationExact}) to hold, it suffices that the approximate equality (\ref{EOMReduced}) hold to within a margin equal to $\frac{\delta_{emp}}{T_{emp}}$. In general, (\ref{EOMReduced}) will hold only for $x_{l}$ in some restricted subset $d \subset S_{l}$. Once the flow  $x_{l}(\tau)$ carries the evolution out of $d$, the approximation in (\ref{EOMReduced}), parametrized by the empirical margins $\delta_{emp}$ and $T_{emp}$, will cease to hold. 
\footnote{It is worth emphasizing here that relation (\ref{EOMReduced}) bears a close resemblance in certain respects to the requirements for reduction proposed by Ernest Nagel and Kenneth Schaffner, for whom reduction required that it be possible to logically deduce approximate versions of the laws of the reduced theory from those of the reducing theory via the use of ``bridge laws" \cite{NagelSS}, \cite{SchaffnerNagRed}. The approach here is in some ways similar in spirit to Nagel/Schaffner approach in showing that, by virtue, of the low-level model's equations of motion, $B(x_{l}))$ approximately satisfies the high-level model's equations of motion. However, unlike their approach, the requirements for reduction here are formalized mathematically within the specific context of group actions over state space manifolds. Moreover, reduction here concerns relations between two specific models of a single fixed system, rather than between entire theories as in the Nagel/Schaffner approach. For recent discussion of Nagel and Schaffer's approach to reduction, see \cite{dizadji2010s}, \cite{SchaffnerNagRed}, \cite{butterfield2011emergence}. 
}
Applying the Chain Rule to (\ref{EOMReduced}) and substituting $V_l^{\mu}(x_l)$ for $\frac{d x_l^{\mu}}{d \tau}$ using the low-level equations of motion $\frac{d x_l}{d \tau} = V_l(x_l)$, one can easily check that (\ref{EOMReduced}) and (\ref{FlowCommutation}) will be satisfied if the push forward of $V_{l}$ under $B$, evaluated at any $x_{l} \in d$, is approximately equal to $V_{h}$ evaluated at $B(x_{l})$ (see Fig. \ref{VectorFields}):

\begin{equation} \label{PushForward}
\frac{\partial B^{\mu}}{\partial x_{l}^{\alpha}}\bigg|_{x_{l}} \ V_{l}^{\alpha}\big|_{x_{l}} \approx V_{h}^{\mu} \big|_{B(x_{l})} \ \ \ \ \ \  \text{for} \ x_{l} \in d. 
\end{equation}

\noindent In fact, one may take the condition (\ref{PushForward}), where $\approx$ holds within $\frac{\delta_{emp}}{T_{emp}}$, to \textit{define} the domain $d \subset S_{l}$. That is, one may define $d$ as the set of states for which (\ref{PushForward}) holds within the required margin of approximation. 
Note that all reference to the flow parameter $\tau$ is removed from this formulation of the matching condition between the dynamical flows prescribed by the different models, so that the requirements for reduction between the models only concern the relationship between the vector fields that generate the dynamical state space flows of the two models. 

\begin{figure}
\center{\includegraphics[width=.6 \linewidth]{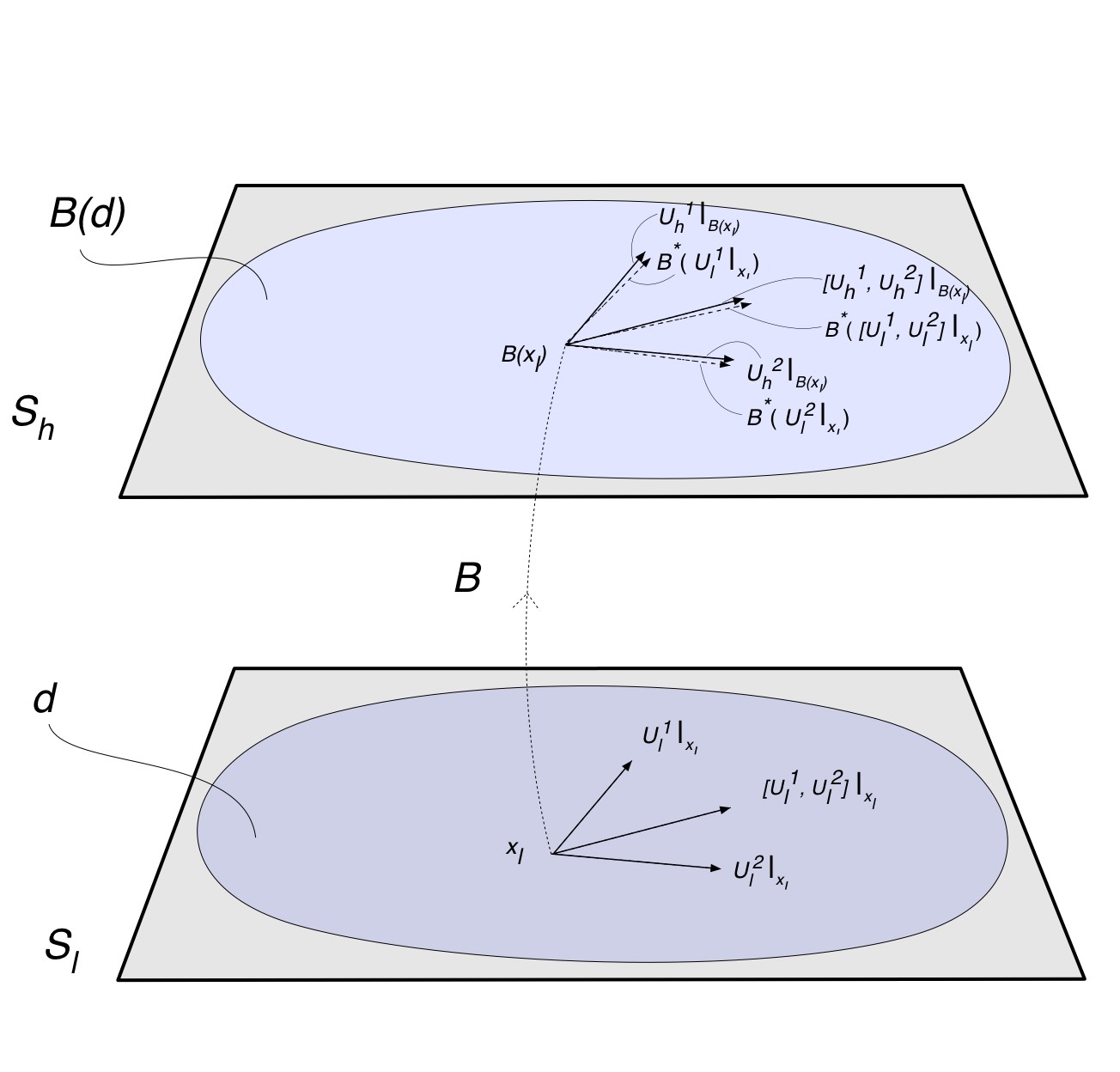}}
\caption{If the high-level dynamical vector field $V_{h}$ evaluated at $B(x_{l})$ is approximately equal to the push-forward of the low-level vector field $V_{l}$ (dotted arrows in upper state space) evaluated at $x_{l}$, for all $x_{l} \in d$, then the trajectory induced on $S_{h}$ through $B$ by the integral curves of $V_{l}$ in $d$ will approximate the integral curves of $V_{h}$ in the image domain $B(d) \subset S_{h}$. 
}
\label{VFieldComposition}
\end{figure}

In fact, it sometimes happens that the relationship (\ref{PushForward}) extends beyond the particular vector fields $V_{h}$ and $V_{l}$ that generate the models' dynamical evolution to include the full algebra of physical symmetry generators 
\footnote{Physical symmetries are understood here as those that map between physically distinct states when the symmetries are interpreted as active transformations, while  gauge symmetries only map between redundant representations of a single physical state and never between physically distinct states.  
}
of these models. In such cases, if $\{V_{h}, U_{h}^{1}, ... , U_{h}^{n} \}$ is a basis for the algebra of vector field generators over $S_{h}$ of $M_{h}$'s physical symmetries, then there exist corresponding $\{V_{l}, U_{l}^{1}, ... , U_{l}^{n} \}$ in the algebra of vector field generators over $S_{l}$ of $M_{l}$'s physical symmetries such that $\frac{\partial B^{\mu}}{\partial x^{\alpha}}|_{x_{l}} U_{l}^{i, \alpha}|_{x_{l}} \approx U_{h}^{i, \mu}|_{B(x_{l})}$ for $x_{l} \in d$, and such that the push forward mapping approximately preserves the Lie algebra structure over $d$:

\begin{align} \label{BridgeFunctionSymmetries}
\frac{\partial B^{\mu}}{\partial x_{l}^{\alpha}}\bigg|_{x_{l}} [V_{l}, U_{l}^{i}]^{\alpha} \big|_{x_l} & \approx [V_{h}, U_{h}^{i}]^{\mu} \big|_{B(x_{l})} \ \ \ \ \ \  \text{for} \ x_{l} \in d \ \ \ \text{for all} \ 1 \leq i \leq n \nonumber \\
\frac{\partial B^{\mu}}{\partial x_{l}^{\alpha}}\bigg|_{x_{l}} [U_{l}^{i}, U_{l}^{j}]^{\alpha} \big|_{x_l} & \approx [U_{h}^{i}, U_{h}^{j}]^{\mu} \big|_{B(x_{l})} \ \ \ \ \ \  \text{for} \ x_{l} \in d \ \ \ \text{for all} \  1 \leq i,j \leq n;
\end{align}

\noindent  see Fig. \ref{VFieldComposition}. The margins of error in the approximate equalities are set to ensure that the generated flows agree within the required empirically determined margin of error. 
For explicit demonstration of this claim with regard to the relationship between unitary and canonical group actions over quantum and classical state spaces (respectively), see \cite{Rosaler2018}. Like the well-known analysis of Inonu and Wigner concerning group contractions, this relationship reveals a particular type of mathematical connection between the symmetry groups of different theories \cite{inonu1953contraction}. However, it is distinct from the analysis of Wigner and Inonu in that on the current approach, the dovetailing between group actions is restricted to low-level group actions operating within a certain subset $d$ of the reducing model's state space; on Inonu and Wigner's approach, there is no such state space restriction, or mention of the need for a mapping $B$ that identifies the low-level model's proxy for $x_{h}$. 

\subsection{Examples}

Let us now examine several examples of the general type of relationship between state space models just described. These examples are specially chosen to illustrate the main claims of the next two sections, concerning the manner in which and conditions under which distinct inter-model reductions may be composed, and a set of consistency constraints that follow when a single reduction may be effected via distinct composite reductions differing in their intermediate layer of description. 
\footnote{Further examples of this relationship may be found in \cite{Rosaler2018}, \cite{rosaler2015local}, and \cite{RosalerThesis}. 
}
We present each example first by specifying state spaces and equations of motion of the models involved and then detailing the form of the function $B$, the relevant state space domain $d \subset S_{l}$, and the specific form of (\ref{EOMReduced}), which in each case can be shown to hold using the low-level equations of motion. We consider all models as alternative descriptions of the same physical system, consisting of a massive charged particle (e.g., proton) in a background electromagnetic potential. 
These examples for the most part serve to formulate known results within the general framework just presented. However, in the following sections, we will begin to see the usefulness of this framework both in clarifying the nature of the relationship between the quantum-to-classical and relativistic-to-non-relativistic transitions, as well as the relationship between speculative theories of new physics and currently established theories. It will also serve to formalize the intuition that different reductions can be composed to form new reductions, and reveal a particular mathematical sense in which different sorts of theoretical transition (e.g., classical-to-quantum and non-relativistic-to-relativistic) may commute with each other.

\subsubsection{Newtonian Mechanics (NM)/ Quantum Mechanics (QM)}

Both Newtonian (i.e., non-relativistic classical) and quantum mechanical models can be used within a certain regime to accurately describe the evolution of a charge particle such as a proton in a background electric field.
\footnote{Note that, despite their subatomic nature, it is the \textit{classical} Lorentz Force Law that is used to guide the motions of charged particle beams in an accelerator.
}
The state space of the classical model, $S_{h} = \Gamma$, is the classical one-particle phase space $\Gamma$; its equations of motion are Hamilton's equations parametrized by the proton mass $m$, charge $q$, electrostatic potential $V$ and static magnetic vector potential $\vec{A}$. The state space of the quantum model, $S_{l} = \mathcal{H}$ is the one-particle Hilbert space (let us focus on the position degree of freedom, ignoring spin); its equation of motion is Schrodinger's equation, parametrized by the same $m$, $q$, $V(\vec{x})$, $\vec{A}(\vec{x})$.
\footnote{Assuming that effects of decoherence can be ignored - which is realistic for such small systems  - the dynamics of the charged particle can be modeled by a purely unitary dynamics to a good approximation. 
}
The bridge function $B: \mathcal{H} \rightarrow \Gamma$ and domain $d \subset \mathcal{H}$ in this case are given by:


\begin{align*}
&B( |\psi \rangle) = \left(\langle \psi | \hat{\vec{x}} | \psi \rangle,  \langle \psi | \hat{\vec{p}} | \psi \rangle \right) \\
&d = \{ \text{``narrow wave packet states"} \}
\end{align*}

\noindent where the wave packets in $d$ are narrow in the specific  sense that their position space widths are small by comparison with the characteristic length scale on which the potentials $V(\vec{x})$ and $\vec{A}(\vec{x})$ vary. By the generalization of Ehrenfest's Theorem for a charge in a static background electromagnetic field 
\footnote{See, e.g., \cite{franklin2016classical} for proof and discussion of this result.
}
with potentials $V$ and $\vec{A}$, it follows from Schrodinger's equation that $| \psi \rangle \in d$, 

\begin{align} \label{EhrenfestApproxNR}
\frac{d \langle \hat{\vec{p}} \rangle}{dt} & \approx q \vec{E}(\langle \hat{\vec{x}} \rangle) + q \frac{d \langle \hat{\vec{x}} \rangle }{dt} \times  \vec{B}( \langle \hat{\vec{x}} \rangle) \nonumber \\
\frac{d  \langle \hat{\vec{x}} \rangle}{dt} & \approx \frac{1}{m} \left(\langle \hat{\vec{p}} \rangle - q \vec{A}\left( \langle \hat{\vec{x}} \rangle \right)\right) ; 
\end{align}

\noindent where $\vec{E}(\vec{x}) = - \nabla V(\vec{x}) - \frac{1}{c} \frac{\partial \vec{A}(\vec{x})}{\partial t} = - \nabla V(\vec{x})$ and  $ \vec{B}(\vec{x}) = \nabla \times \vec{A}(\vec{x})$. The terms on the right-hand side can be understood as components of the Hamiltonian vector field that generates dynamical flows over classical phase space. Thus, $B( |\psi \rangle)$ approximately satisfies the classical Hamilton equations, and so satisfies (\ref{EOMReduced}). This will only hold over the timescales over which $| \psi \rangle$ remains in $d$ - i.e. the timescales over which the wave packet remains sufficiently narrowly peaked in position. This timescale will generally be longer for larger values of $m$, which reduce the rate of wave packet spreading under the Schrodinger evolution.
\footnote{See, e.g., \cite{RosalerThesis}, Ch. 2 for further discussion of this well-known result. 
}
The neglected errors in the approximate equalities $\approx$ are proportional to the position and momentum widths of the quantum state $| \psi \rangle$.

\subsubsection{Quantum Mechanics (QM)/ Relativistic Quantum Mechanics (RQM)}

Models of both non-relativistic and relativistic quantum mechanics can be used to accurately describe the behavior of a low momentum ($<< mc$), but not necessarily localized, charge such as a proton. The state space of the quantum model, $S_{h} = \mathcal{H}_{Pauli}$, now explicitly including spin, is the Pauli Hilbert space of 2-spinors; the equation of motion of such a model is the Pauli equation for non-relativistic spin-1/2 particles, with parameter values $m$ and $q$, and background potentials $V(\vec{x})$, $\vec{A}(\vec{x})$. The state space of the RQM model, $S_{l} = \mathcal{H}_{Dirac}$, is the Dirac Hilbert space of 4-spinors; the model's equation of motion is the Dirac equation with parameters $q$, $m$, $V$, and $\vec{A}$ as above. The bridge function $B: \mathcal{H}_{Dirac} \rightarrow \mathcal{H}_{Pauli}$ and domain $d \subset \mathcal{H}_{Dirac}$ are given by


\begin{align*}
& B^{\alpha}[\psi^{a}(\vec{x},t)] = e^{imt}P^{\alpha}_{a} \psi^{a}(\vec{x},t) \\
& d = \{ `\text{``low-momentum, positive energy 4-spinors"} \}
\end{align*}

\noindent where $\psi^{a}(\vec{x},t)$ is the 4-spinor wave function, and the operator $P^{\alpha}_{a}$ projects 4-spinors onto their upper two components. It is well known that for low-momentum four-spinors in the Dirac representation - i.e., for $\psi^{a}(\vec{x}) \in d$ - the upper two components approximately satisfy the Pauli equation:

\begin{footnotesize}
\begin{align} \label{Pauli}
 i\hbar \frac{\partial}{\partial t} \left( e^{imt} P_{a}^{\alpha}  \psi^{a}(\vec{x},t) \right)  \approx   \left\{  \frac{1}{2 m}  \left[ \sigma \cdot \left( -i \hbar \nabla  - q \vec{A}(\vec{x})   \right)\right]^{2} + q V(\vec{x}) \right\}^{\alpha \beta}    \left( e^{imt} P_{a}^{\beta}   \psi^{a }(\vec{x},t) \right)\ .
\end{align}
\end{footnotesize}

\noindent This shows that (\ref{EOMReduced}) is satisfied in this case; see, e.g., \cite{DineNRDirac} for proof of this claim. This approximate equality will only hold for as long as the function $\psi^{a}(\vec{x},t) $ remains in the subset $d$ of low-momentum states. The neglected error in the approximation $\approx$ is proportional to $\frac{\mu^{4}}{m^{4}}$, where $\mu$ is the upper limit of modes in the momentum-space expansion of $\psi^{a}(\vec{x},t) $. 

\subsubsection{Special Relativity (SR)/ Relativistic Quantum Mechanics (RQM)}

Both relativistic classical and relativistic quantum models may be used to describe a charge with any kinetic energy, as long as its quantum mechanical wave packet is not too spread out. The state space of the classical model is the relativistic classical phase space $S_{h} = \Gamma_{rel}$; the equations of motion are the relativistic Hamilton equations with parameters $m$, $q$, $V(\vec{x})$, and $\vec{A}(\vec{x})$ as above. The state space of the quantum model is the Dirac Hilbert space $S_{l} = \mathcal{H}_{Dirac}$; the equations of motion are as in the previous example. The bridge function $ B: \mathcal{H}_{Dirac} \rightarrow \Gamma_{rel}$ and domain $d \subset \mathcal{H}_{Dirac}$ are given by

\begin{align*}
&B[\psi^{a}(\vec{x})] = \left( \int d^{3} x \ \vec{x} \ \psi^{a \dagger}(\vec{x})  \psi^{a}(\vec{x}) ,    \int d^{3} x \ \psi^{a \dagger}(\vec{x})  (-i \hbar \nabla)  \psi^{a}(\vec{x})  \right) = \left( \langle \hat{\vec{x}} \rangle, \langle \hat{\vec{p}} \rangle \right)\\
& d = \{ \text{``narrow 4-spinor wave packets"} \}. 
\end{align*}

\noindent $B$ maps a 4-spinor into the corresponding expectation values of 3-position and 3-momentum, and $d$ consists of Dirac spinor fields that are narrowly peaked in position and momentum (although the degree to which they may be simultaneously peaked in both is of course constrained by the uncertainty principle). In this case, one can prove a relativistic analogue of Ehrenfest's Theorem, from which it follows that expectation values approximately satisfy the relativistic Lorentz Force Law:

\begin{align}\label{EhrenfestApproxSR}
\frac{d}{dt} \langle \hat{\vec{p}} \rangle & \approx q \vec{E}(\langle \hat{\vec{x}} \rangle) + q  \frac{d\langle \hat{\vec{x}} \rangle} {dt} \times  \vec{B}(\langle \hat{\vec{x}} \rangle) \nonumber \\
\frac{d}{dt} \langle \hat{\vec{x}} \rangle & \approx \frac{1}{m \gamma(\langle \hat{\vec{p}} \rangle ) } \langle \hat{\vec{p}} \rangle
\end{align}

\noindent for $\psi^{a}(x) \in d$; see \cite{RosalerThesis}, Ch.4 for proof of this claim. Thus, (\ref{EOMReduced}) holds in this case as well, but only for as long the wave packet remains sufficiently narrow in both position and momentum (within the constraints of the uncertainty principle). The neglected errors in the approximation $\approx$ are proportional to the position and momentum space widths of the Dirac wave packet.

\subsubsection{Newtonian Mechanics (NM)/ Special Relativity (SR)}

A slow-moving charge may be described either by a model of Newtonian mechanics or of classical relativistic mechanics. As we have seen, the state space of the classical model is the classical one-particle phase space $S_{h}=\Gamma$; its equations of motion are Hamilton's equations parametrized by the mass $m$ and charge $q$ and potententials $V(\vec{x})$, $\vec{A}(\vec{x})$. The state space of the classical model is the relativistic classical phase space $S_{h} = \Gamma_{rel}$; the equations of motion are the relativistic Hamilton equations with parameters $m$, $q$, $V(\vec{x})$, $\vec{A}(\vec{x})$ as above. The bridge function $B: \Gamma_{rel} \rightarrow \Gamma$ and domain $d \subset \Gamma_{rel}$ are given simply by

\begin{align*}
& B(\vec{x}, \vec{p}) \equiv (\vec{x}, \vec{p}). \\
& d = \{ (\vec{x}, \vec{p}) \in \Gamma_{rel} \big| \ |\vec{v}|  << c \}
\end{align*} 

\noindent where $\vec{v} = \frac{c \vec{p} - q \vec{A(\vec{x})}}{\sqrt{(\vec{p} - \frac{q}{c} \vec{A}(\vec{x}))^{2} + m^{2} c^{2}}}$. The relation (\ref{EOMReduced})  in this case takes the form,

\begin{align} \label{NRApprox}
\frac{d \vec{p} }{dt} & = q \vec{E}(\vec{x}) + q \frac{d \vec{x} }{dt} \times  \vec{B}( \vec{x}) \nonumber \\
\frac{d \vec{x} }{dt} & \approx \frac{1}{m} \left(\vec{p} - q \vec{A}\left( \vec{x} \right)\right) ; 
\end{align}


\noindent and follows straightforwardly from expansion in powers of $\frac{v}{c}$. As always, this approximation holds only for as long as the low-level state remains in the subset $d$ - that is, as long as $|\vec{v}|  << c$. The neglected terms in the approximate equality $\approx$ are proportional to $\frac{|\vec{v}|^3}{c^3}$.



\subsection{A Note on Gauge Theories}

The framework for reduction described above assumes that the reduced and reducing models both can be formulated in terms of deterministic flows generated by some dynamical vector fields over their respective state spaces. While these conditions are met by many pairs of models between which reduction holds, models of gauge theories do not satisfy these requirements, since models of gauge theories do not prescribe a deterministic evolution for the gauge fields unless gauge has been fixed. But the choice of gauge is purely a matter of convention and has no physical import. In recovering the empirical successes of a model associated with some gauge theory, it is only necessary to approximate the gauge invariant features of the state evolution by some induced evolution $B(x_{l}(\tau))$. In the case of classical electrodynamics, this would consist only of the transverse components of the electromagnetic 4-potential $A^{\mu}(x)$. In quantized gauge theories, one likewise should only demand to recover gauge-invariant features of the state evolution, although the precise manner of doing so depends on the specific method of quantization that is used to define the theory.

\section{Compound Reduction and Overlapping State Space Domains} \label{ReductionCommute}

Intuitively, it is natural to expect that reduction should be transitive: if, relative to some physical system $K$, model $M_{1}$ reduces to model $M_{2}$ and model $M_{2}$ reduces to model $M_{3}$, then model $M_{1}$ should reduce directly to model $M_{3}$. This intuition can be formalized straightforwardly within the general framework described in the previous section. 

Denote the bridge function and state space domain for the reduction of $M_{1}$ to $M_{2}$ (abbreviated $1 \rightarrow 2$) respectively as $B_{2}^{1}$ and $d_{2}^{1}$, and the bridge function and state space domain for the reduction of $M_{2}$ to $M_{3}$ (abbreviated $2 \rightarrow 3$) respectively as $B_{3}^{2}$ and $d_{3}^{2}$. 
Then the bridge function and state space domain of the direct reduction of $M_{1}$ to $M_{3}$ (abbreviated $1 \rightarrow 3$) are given respectively by

\begin{align}  \label{CompoundReductionGeneral}
B^{1}_{3}(x_{3}) &= B^{1}_{2}(B^{2}_{3}(x_{3})) \nonumber \\
d_{3}^{1} &= d_{3}^{2} \cap B_{3}^{2,-1}(d_{2}^{1}), 
\end{align}

\noindent where $B_{3}^{2,-1}(d_{2}^{1})$ is the inverse image of $d_{2}^{1}$ under $B_{3}^{2}$. Reduction of $M_{1}$ to $M_{3}$ occurs only if the set $d_{3}^{1}$ is non-empty - that is, only if the image domain $B_{3}^{2}(d_{3}^{2})$ of the low-level, 2-to-3 reduction overlaps with the state space domain $d_{2}^{1}$ of the high-level, 1-to-2 reduction. The set $d_{3}^{1} = d_{3}^{2} \cap B_{3}^{2,-1}(d_{2}^{1})$ is then just the subset of the state space domain of the $2 \rightarrow 3$ reduction whose image under $B_{3}^{2}$ lies in the state space domain of the $1 \rightarrow 2$ reduction. Reduction of $M_{1}$ to $M_{3}$ requires that the quantity $B_{3}^{1}(x_{3}(\tau))$, whose behavior is governed by the dynamics of $M_{3}$, approximately satisfy the equations of motion of $M_{1}$ when $x_{3}(\tau) \in d_{3}^{1} = d_{3}^{2} \cap B_{3}^{2,-1}(d_{2}^{1})$:

\begin{align} \label{EOMReducedCompose}
\frac{d}{d \tau} B^{1, \mu}_{2}(B^{2}_{3}(x_{3}(\tau))) &\approx V_{1}^{\mu}\big|_{B^{1}_{2}(B^{2}_{3}(x_{3}(\tau)))} \ \ \ \text{for} \ \ \ x_{3} \in d_{3}^{2} \cap B_{3}^{2,-1}(d_{2}^{1}),
\end{align}

\noindent where on the left-hand-side we have made use of the substitution $B^{1}_{3}(x_{3}) = B^{1}_{2}(B^{2}_{3}(x_{3}))$. The approximate equality $\approx$ is constrained by the parameters $\delta_{emp}^{1}$ and $T_{emp}^{1}$ characterizing $M_{1}$'s accuracy in describing the physical degrees of freedom $K$.  

The reduction $1 \rightarrow 3$ follows deductively from the reductions $1 \rightarrow 2$ and $2 \rightarrow 3$. Although (\ref{EOMReducedCompose}) is not difficult (just somewhat tedious) to prove rigorously using exact inequalities of the form (\ref{FlowCommutationExact}), we can see intuitively how (\ref{EOMReducedCompose}) follows from the reductions $1 \rightarrow 2$ and $2 \rightarrow 3$ using approximate equalities. In the reduction $1 \rightarrow 2$, the fact that $x_{2}(\tau)$ exactly satisfies $M_{2}$'s equations of motion implies that the quantity $B^{1}_{2}(x_{2}(\tau))$ approximately satisfies the $M_{1}$ equations of motion for $x_{2}(\tau) \in d_{2}^{1}$. By contrast with $x_{2}(\tau)$, the quantity $x^{'}_{2}(\tau) \equiv B^{2}_{3}(x_{3}(\tau))$ only \textit{approximately} satisfies the $M_{2}$ equations, for $x_{3}(\tau) \in d_{3}^{2}$. Given the continuity of the bridge functions (which in practice always holds), the fact that $B^{1}_{2}(x_{2}(\tau))$ approximately satisfies the $M_{1}$ equations of motion should not be significantly altered when $x_{2}(\tau)$ is adjusted to be only an approximate, rather than exact, solution to the $M_{2}$ equations. Thus, $B_{2}^{1}(x^{'}_{2}(\tau))$, which is equal to $B^{1}_{2}(B^{2}_{3}(x_{3}(\tau)))$, also approximately satisfies the $M_{1}$ equations of motion, as we wished to show. 
\footnote{See \cite{RosalerThesis}, Ch.1 for a more formal proof of this claim.
}
Using the Chain Rule, one can eliminate reference to the flow parameter $\tau$ in (\ref{EOMReducedCompose})  via a relation of the type (\ref{PushForward}):

\begin{align} \label{PushForwardCompose}
\frac{\partial B^{1, \mu}_{2}}{\partial x_{2}^{\alpha}}\bigg|_{B_{3}^{2}(x_{3})} \frac{\partial B^{2, \alpha}_{3}}{\partial x_{3}^{\beta}}\bigg|_{x_{3}} \ V_{3}^{\beta}\big|_{x_{l}} \approx V_{1}^{\mu}\big|_{B^{1}_{2}(B^{2}_{3}(x_{3}(\tau)))} \ \ \ \text{for} \ \ \ x_{3} \in d_{3}^{2} \cap B_{3}^{2,-1}(d_{2}^{1}),
\end{align}

\noindent where $\frac{\partial B^{1, \alpha}_{3}}{\partial x_{3}^{\nu}}\big|_{x_{3}}  = \frac{\partial B^{1, \alpha}_{2}}{\partial x_{2}^{\mu}}\big|_{B_{3}^{2}(x_{3})} \frac{\partial B^{2,\mu}_{3}}{\partial x_{3}^{\nu}}\big|_{x_{3}} $. The requirements for reduction are then formulated exclusively in terms of the vector fields that generate the dynamical evolution over the models' state spaces. One can further extend this push-forward relationship to draw a direct connection between the different algebras of physical symmetries over the various state spaces involved a composite reduction, via relations of the form (\ref{BridgeFunctionSymmetries}). 

\subsection{Consistency Requirements between Alternative Composite Reductions}
 
In certain cases, a single direct reduction $1 \rightarrow 3$ can be effected via either of two intermediate models $M_{2a}$ or $M_{2b}$, via either of the composite reductions $1 \rightarrow 2a \rightarrow 3$ or $1 \rightarrow 2b \rightarrow 3$. We will see an example of this in the following subsection, where the reduction of a non-relativistic classical model to a model of relativistic quantum mechanics may be effected either via a model of non-relativistic quantum mechanics or a model of relativistic classical mechanics. Such cases correspond roughly to the commutation of limits along a single face of the Bronstein cube, although the Bronstein cube makes no mention of the need for bridge functions or restricted state space domains. In cases where is possible to effect the reduction $1 \rightarrow 3$ either via the path $1 \rightarrow 2a \rightarrow 3$ or $1 \rightarrow 2b \rightarrow 3$, we are free to consider the models $M_{1}$ to $M_{3}$ without regard to the possibility of intermediate models $M_{2}$ between them. For a given physical system $K$, there is some fact of the matter as to which quantities in $M_{3}$ serve to instantiate the physical degrees of freedom described by $M_{1}$ in cases where $M_{1}$ is empirically accurate, and which subset of states in $S_{3}$ allow for the approximate description of these degrees of freedom by a Newtonian model. It is then merely a matter of how we choose to describe the system, rather than a matter of physical fact, whether we view the $1 \rightarrow 3$ reduction as the composition of the $1 \rightarrow 2a$ and $2a \rightarrow 3$ reductions, or as the composition of the $1 \rightarrow 2b$ and $2b \rightarrow 3$ reductions. This suggests that the bridge function $B_{3}^{1}(x_{3})$ and state space domain $d_{3}^{1}$ should be independent of which of these ``reduction paths" is used:

\begin{align} \label{Compatibility}
\Aboxed{B^{1}_{3}(x_{3}) & \apeq B^{1}_{2a}(B^{2a}_{3}(x_{3})) \approx B^{1}_{2b}(B^{2b}_{3}(x_{3})) \ \text{for} \ x_{3} \in d_{3}^{1}} \\
& \nonumber \\
\Aboxed{d^{1}_{3} &\approx d_{3}^{2a} \cap B_{3}^{2a,-1}(d_{2a}^{1}) \approx d_{3}^{2b} \cap B_{3}^{2b,-1}(d_{2b}^{1})} \label{Compatibility2}
\end{align}

\noindent 
The approximate equivalence $\apeq $ indicates that there is a small ``halo" of non-uniqueness to the bridge function $B^{1}_{3}(x_{3})$, corresponding to a small neighborhood of functions all of which approximately instantiate the same high-level (i.e., $M_{1}$) behavior within the empirical margin of approximation $\delta_{emp}^{1}$. There exists a corresponding blurring around the edges of the state-space domain $d^{1}_{3}$ resulting from the margin $\delta_{emp}^{1}$ and the slight ambiguity in $B^{1}_{3}(x_{3})$. We will see this in more detail when we discuss an example in the next subsection. It is also worth emphasizing that these consistency requirements can naturally be extended to cases where more than two reduction paths are available between the reduced and reducing models, and in which the individual reduction paths chain together more than three distinct models.


\subsection{Commutation of Quantum-to-Classical and Relativistic-to-Non-Relativistic Transitions}

Let us check the conditions (\ref{Compatibility}) and (\ref{Compatibility2}) explicitly for the case of the reduction $NM \rightarrow RQM$ in the description of a slow-moving heavy charge, which may be effected via either of the composite reductions $NM \rightarrow QM \rightarrow RQM$ or $NM \rightarrow SR \rightarrow RQM$. 

Considering first the path $NM \rightarrow QM \rightarrow RQM$, let us check (\ref{CompoundReductionGeneral}), recalling the bridge functions for the $NM \rightarrow QM$ and $QM \rightarrow RQM$ reductions in Section \ref{BridgeFunctions}, and the fact that $x_{RQM}$ is specified by the 4-spinor wave function $ \psi^{a}(\vec{x})$. Composing these functions yields,

\begin{align}
 B^{NM}_{RQM}(x_{RQM}) &= B^{NM}_{QM}(B^{QM}_{RQM}(x_{RQM})) \nonumber \\
 &=  \left(\sum_{a=1}^{2} \int d^{3} x \ \vec{x} \ \psi^{a \dagger}(\vec{x})  \psi^{a}(\vec{x}), \sum_{a=1}^{2} \int d^{3} x \ \psi^{a \dagger}(\vec{x}) (-i \vec{\nabla})  \psi^{a}(\vec{x})   \right),
 \end{align}
 
 \noindent where we emphasize that the sum in the expectation values is only over the upper two 4-spinor components. Likewise, recalling the state space domains of the reductions $NM \rightarrow QM$ and $QM \rightarrow RQM$ from Section \ref{BridgeFunctions}, we can identify the state space domain $d_{RQM}^{NM}$ for the composite reduction $NM \rightarrow QM \rightarrow RQM$ using the relation (\ref{CompoundReductionGeneral}):
 
\begin{align}
d_{RQM}^{NM} &= d_{RQM}^{QM} \cap B_{RQM}^{QM,-1}(d_{QM}^{NM}) \nonumber \\
&= d_{RQM}^{QM} \cap B_{RQM}^{QM,-1}( \{ \text{``narrow wave packets"} \} )  \nonumber \\
&= \{ \text{ ``low-momentum, positive energy 4-spinors" }\} \cap \{ \text{``narrowly peaked positive energy 4-spinors"} \} \nonumber \\
&= \{\text{ ``narrowly peaked, low-momentum positive energy 4-spinors"} \}
\end{align}

\noindent In the second line, we have made use of the fact that the inverse image of $d_{QM}^{NM}$, the set of narrowly peaked non-relativistic wave functions, under $B_{RQM}^{QM}$, which projects 4-spinors onto their upper two components,  is the set of narrowly peaked 4-spinor wave packets. In the third line, there is some tension between the requirement that wave packets be restricted to low-momentum Fourier components and that they be narrowly peaked in position, since narrowly peaked wave packets require high-momentum modes. However, assuming that the external potential fields do not vary too sharply, the intersection between the set of narrow wave packets and the set of low-momentum wave packets may still be sizable (depending on the precise margins one uses to define ``narrow" and ``low-momentum"), and supports a robust approximation of the Newtonian evolution by the evolution induced by the RQM model. That is, wave packets can be simultaneously sufficiently narrow and sufficiently non-relativistic to support both the classical and non-relativistic approximations in tandem. 

Now consider the form of (\ref{CompoundReductionGeneral}) for the alternative reduction path $NM \rightarrow SR \rightarrow RQM$:

\begin{align} \label{BridgeCompositeSR}
 B^{NM}_{RQM}(x_{RQM}) &= B^{NM}_{SR}(B^{SR}_{RQM}(x_{RQM})) \nonumber \\
 &=  \left(\sum_{a=1}^{4} \int d^{3} x \ \vec{x} \ \psi^{a \dagger}(\vec{x})  \psi^{a}(\vec{x}), \sum_{a=1}^{4} \int d^{3} x \ \psi^{a \dagger}(\vec{x}) (-i \vec{\nabla})  \psi^{a}(\vec{x})   \right),
 \end{align}
 
 \noindent where we emphasize that the sum is now over all 4-spinor components and not just the upper two. Informally, we can identify the state space domain for the composite reduction $NM \rightarrow SR \rightarrow RQM$ as
 
 \begin{align}
d_{RQM}^{NM} &= d_{RQM}^{SR} \cap B_{RQM}^{SR,-1}(d_{SR}^{NM}) \nonumber \\
&= d_{RQM}^{SR} \cap B_{RQM}^{SR,-1}( \{ \text{``low-momentum phase space points"} \} ) \nonumber \\
&= \{ \text{ ``narrowly peaked, positive energy 4-spinors" }\} \cap \{ \text{``low-momentum, positive energy 4-spinors"} \} \nonumber \\
&= \{\text{ ``narrowly peaked, low-momentum positive energy 4-spinors"} \}
\end{align}

\noindent In the second line, we have made use of the fact that the inverse image of $d^{NM}_{SR}$, the set of low-velocity phase space points, under $B^{SR}_{RQM}$, which maps 4-spinor wave functions into expectation values of position and momentum, is the set of low-momentum, positive-energy 4-spinor wave packets. 

\begin{figure}
\center{\includegraphics[width=.8 \linewidth]{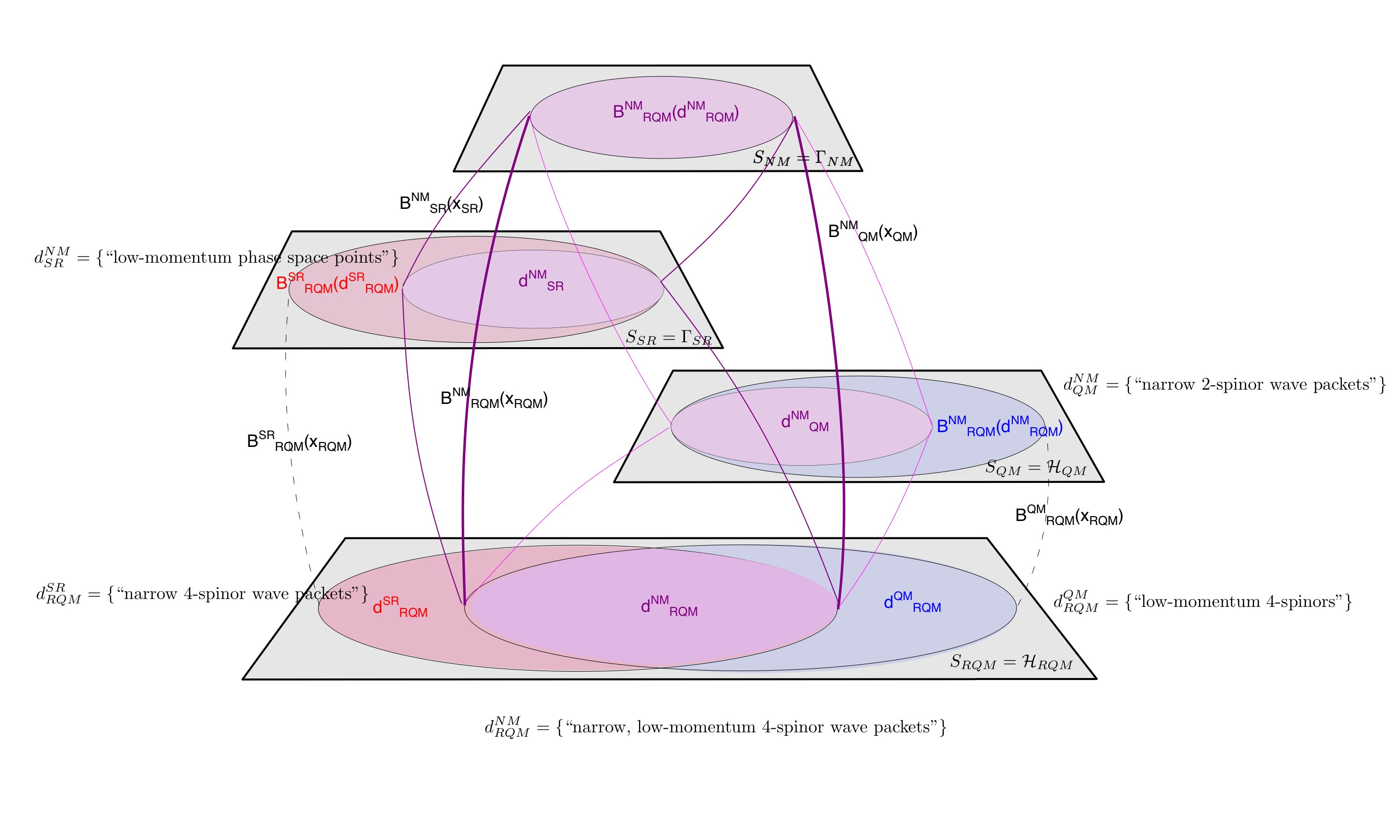}}
\caption{\scriptsize The direct reduction of the Newtonian model $M_{NM}$  of a localized, slow-moving (i.e., non-relativistic) charge to the RQM model $M_{RQM}$ of the same charge may be effected via either of the compound reductions $NM \rightarrow QM \rightarrow RQM$ or $NM \rightarrow SR \rightarrow RQM$. The bridge function $B^{NM}_{RQM}$ and state space domain $d_{RQM}^{NM}$ characterizing the direct reduction $NM \rightarrow RQM$ are approximately independent of which of these two paths one chooses. This illustrates one concrete sense in which the quantum-to-classical and relativistic-to-non-relativistic transitions commute.
}
\label{QuantRelComposite}
\end{figure}

Thus, we see that the state space domains for the two composite reductions are equal along the two reduction paths, consisting in both cases of narrow, low-momentum, positive energy 4-spinors. Note also that $d^{NM}_{SR} = d^{QM}_{RQM} \cap d^{SR}_{RQM}$, where $d^{QM}_{RQM} $ is the set of low-momentum 4-spinors and $d^{SR}_{RQM} $ is the set of narrowly peaked 4-spinors. That is, the state space domain of the reduction $NM \rightarrow RQM$ is simply the overlap between the state space domains $d^{QM}_{RQM}$ and $d^{SR}_{RQM}$ of the two intermediate models. On the other hand, the composite bridge functions for the two reduction paths, $B^{NM}_{QM}(B^{QM}_{RQM}(x_{RQM}))$ and $B^{NM}_{SR}(B^{SR}_{RQM}(x_{RQM}))$, are not strictly speaking equal. The first composite bridge function only involves a summation over the upper two 4-spinor components while the second involves a summation over all four components. However, over  $d^{NM}_{RQM}$, the composite bridge functions associated with the two reduction paths approximately agree, since the contribution of the $3, 4$ components to the sum in (\ref{BridgeCompositeSR}) is very small over this subset of the RQM state space and so can be neglected. Thus, we see from these two composite reductions that 

\begin{align}
B^{NM}_{QM}(B^{QM}_{RQM}(x_{RQM})) &\approx B^{NM}_{SR}(B^{SR}_{RQM}(x_{RQM})) \\
d_{RQM}^{QM} \cap B_{RQM}^{QM,-1}(d_{QM}^{NM}) &\approx d_{RQM}^{SR} \cap B_{RQM}^{SR,-1}(d_{SR}^{NM}) 
\end{align}

\noindent and therefore that the requirements (\ref{Compatibility}) and (\ref{Compatibility2}) are satisfied in this case (see Figure \ref{QuantRelComposite}). To be fully rigorous, one should show that  the size of the neglected terms in these approximate equalities is smaller than the empirical error bound $\delta_{emp}^{NM}$ within which $M_{NM}$ is known to successfully describe the physical degrees of freedom under consideration. 

The approach to reduction based on the specification of bridge functions $B: S_{l} \rightarrow S_{h}$ and state space domains $d \subset S_{l}$ thus illustrates explicitly one important sense in which the quantum-to-classical and relativistic-to-nonrelativistic transitions \textit{commute}. From the perspective of the state space domains, the state space domain $d^{NM}_{RQM}$ of the reduction $NM \rightarrow RQM$ is the same irrespective of whether one first applies the restriction to non-relativistic states $d^{QM}_{RQM}$ and then to classical states $d^{SR}_{RQM}$, or instead applies these restrictions in the reverse order; this simply reflects the fact that $A \cap B = B \cap A$ for arbitrary sets $A$ and $B$. From the perspective of bridge functions, the specific quantity identified as the RQM proxy for an NM phase space space point is independent of whether one first applies the map from to the non-relativistic state space and then from there to the classical non-relativistic state space, or instead first applies the map to the relativistic classical phase space and then to the non-relativistic classical phase space. Within the margin of approximation and timescales for which the high-level Newtonian model successfully describes the position and momentum of the physical charge in question, the bridge function $B_{RQM}^{NM}: \mathcal{H}_{RQM} \rightarrow \Gamma_{NM}$ and state space domain $d_{RQM}^{NM} \subset \mathcal{H}_{RQM}$ characterizing the direct reduction of $NM \rightarrow RQM$ are independent of whether the reduction is performed via the path $NM \rightarrow QM \rightarrow RQM$ or the path $NM \rightarrow SR \rightarrow RQM$. 

In either case, the quantity $B_{RQM}^{NM}[\psi^{a}(\vec(x))]$, understood as a composition of bridge functions associated with the component reductions along \textit{either} path, can be shown to approximately satisfy the high-level non-relativistic classical Hamilton equations for $(\vec{x}, \vec{p}) \in \Gamma$ when $ [\psi^{a}(\vec{x})] \in d^{NM}_{RQM}$:

{\scriptsize
\begin{align}
\frac{d}{dt}  \bigg( \sum_{a=1}^{4 (\text{or 2})} \int d^{3} x  \  \psi^{a \dagger}(\vec{x}, t)  (-i \hbar \nabla - q \vec{A}(\vec{x}))    \psi^{a}(\vec{x}, t) &  \bigg) \approx \vec{E} \left( \sum_{a=1}^{4 (\text{or 2})}  \int d^{3} x \ \vec{x} \ \psi^{a \dagger}(\vec{x}, t)   \psi^{a}(\vec{x}, t)  \right) \nonumber \\
 +  \frac{1}{m} \bigg(\sum_{a=1}^{4 (\text{or 2})}  \int d^{3} x & \ \psi^{a \dagger}(\vec{x}, t) (-i \hbar \nabla - q \vec{A}(\vec{x}))   \psi^{a}(\vec{x}, t)  \bigg)  \times \vec{B} \left(\sum_{a=1}^{4 (\text{or 2})}  \int d^{3} x \ \vec{x} \ \psi^{a \dagger}(\vec{x}, t)   \psi^{a}(\vec{x}, t)  \right)   \\
\frac{d}{dt}  \left(\sum_{a=1}^{4 (\text{or 2})} \int d^{3} x \ \vec{x} \ \psi^{a \dagger}(\vec{x}, t)  \psi^{a}(\vec{x}, t)  \right) & \approx  \frac{1}{m} \left( \sum_{a=1}^{4 (\text{or 2})} \int d^{3} x \ \psi^{a \dagger}(\vec{x}, t) (-i \hbar \nabla - q \vec{A}(\vec{x}))   \psi^{a}(\vec{x}, t)  \right) 
\end{align}
}

\noindent where the electric and magnetic fields are defined in terms of the potentials appearing in the Dirac equation by the relations $\vec{E}(\langle \vec{x} \rangle) \equiv -\nabla{V}(\langle \vec{x} \rangle)$ and $\vec{B}(\langle \vec{x} \rangle) \equiv \nabla \times \vec{A} (\langle \vec{x} \rangle) $, with $\langle \vec{x} \rangle =   \int d^{3} x \ \vec{x} \ \psi^{a \dagger}(\vec{x}, t)   \psi^{a}(\vec{x}, t) $. The upper index on the sums in the above expression may be taken as either $2$ or $4$ since the lower two components are negligible over $d^{NM}_{RQM}$. Thus, the relation (\ref{EOMReducedCompose}) is satisfied simultaneously for both paths $NM \rightarrow QM \rightarrow RQM$ and $NM \rightarrow SR \rightarrow RQM$. 

Here, we see that a quantity constructed within the remote and theoretically abstract realm of four-spinors and Dirac matrices instantiates the much more intuitive and familiar behavior of Newtonian mechanics. In principle, reduction of the RQM model to even more fundamental descriptions such as a model of QED could be used to effect a direct reduction of $M_{NM}$ to $M_{QED}$, which in turn would enable us to replace the expression for the Newtonian state in terms of 4-spinor wave functions with an expression in terms of, say, the QED state. 
Through the use of bridge functions and state space domain restrictions, it is possible to describe the familiar, intuitive phenomena more direct to our experience in the abstract theoretical terms needed to characterize phenomena and degrees of freedom far more remote from our direct experience. In principle, it should be possible via these methods (or a suitably generalized version of them within the instantiation-based way of conceptualizing reduction) to embed the phenomena of everyday experience into the more universal, but far more abstract, theoretical framework furnished by some particular model of quantum gravity. Thus, the use of bridge functions and state space domains facilitates more explicit articulation of the the reductionist ideal
according to which the theories and models of physics become progressively more universal, which has driven theoretical progress since Galileo and Kepler and currently motivates the search for a theory of quantum gravity or a ``theory of everything." 
\footnote{Within the recent philosophical literature, the metaphysical basis for and implications of such a reductionist view, including careful analysis of what it means for one theory to be more ``fundamental" than another, is examined  extensively in the work of Ladyman, French, Saatsi, McKenzie, and many others \cite{ladyman2007every}, \cite{french2011defence}, \cite{saatsi2016theoretical}, \cite{mckenzie2017against}. 
} 

\subsection{Making Approximate Equalities Exact in the Limit}

The preceding analysis of the classical and non-relativistic domains of an RQM model kept the constants $\hbar$ and $c$ fixed at their physical values. There it was typically the case that many of the equalities between quantities in a high-level model $M_{h}$ and their instantiations by some low-level model $M_{h}$ were approximate in nature. In this section, we explain how these approximate equalities, which hold in a certain sense ``on the way" to the limit, can be made exact by counterfactually varying the values of the constants $\hbar$ and $c$. 
\footnote{Butterfield employs this distinction, between behavior in the limit, and on the way to the limit, in order to propose a reconciliation between the clashing concepts of reduction and emergence in physics \cite{butterfield2011less}.  
} 

Consider first the non-relativistic approximation to relativistic theories. Before considering the counterfactual limit, it is worth inquiring to what extent the approximate equalities in relations of the form (\ref{FlowCommutation}) and (\ref{EOMReduced}) can be made exact \textit{without} changing the value of $c$. In the reduction  $NM \rightarrow SR$, the approximate equality becomes exact only when $\gamma = 1$ - that is, when the velocity is exactly zero. For any non-zero value of $v$ with $c$ fixed, $\gamma > 1$, so that the approximation cannot be exact. Thus, under these assumptions, for fixed $c$ the non-relativistic approximation cannot hold exactly except when $v=0$. Likewise, in the reduction $QM \rightarrow RQM$, the upper two components of the Dirac 4-spinor satisfy the Pauli equation exactly only in the case of constant 4-spinors that do not possess Fourier modes for which $|\vec{k}| > 0$ - i.e., 4-spinors that are spatially constant. Thus, in both reductions, the approximate equalities become exact only in the highly restricted case of physical states in which the system is at rest or lacks any component with finite momentum. If, on the other hand, one takes the formal, counterfactual limit $c \rightarrow \infty$, then the approximate equalities (\ref{Pauli}) and (\ref{NRApprox}) should become exact for \textit{all} states in $\mathcal{H}_{Dirac}$ and $\Gamma_{rel}$, so that state space domain restrictions are no longer needed to recover approximate (or exact) validity of the high-level equations of motion as applied to $B(x_{l})$. However, it is precisely for this reason that attempts to effect these reductions on a Bronstein-style approach overlook the necessity of restricting to a particular domain $d$ of the physical state space when identifying those quantities in the reducing model that approximately instantiate the regularities of the reduced model, and when circumscribing the circumstances under which they do so.

In the case of quantum-to-classical transitions, we can likewise ask whether the approximate equalities in relations of the form (\ref{FlowCommutation}) and (\ref{EOMReduced}) can be made exact without changing the value of $\hbar$. The requirement that expectation values approximately satisfy classical equations of motion, reflected in (\ref{EhrenfestApproxNR}) and (\ref{EhrenfestApproxSR}), depends on the possibility of wave packets that are simultaneously narrowly peaked both in position and momentum: if they are too narrowly peaked in momentum, they will have a wide spatial spread with respect the characteristic length scale of the background potential, so that Ehrenfest's Theorem no longer implies approximately classical evolutions for expectation values (not to mention the fact that states with wide spatial spread are inherently non-localized and therefore non-classical); on the other hand, if the quantum state is too narrowly peaked in position (such as in the case of a delta function), it will be localized and evolve classically only for the very briefest of instants, so that classicality fails to persist for any extended period of time. Thus, a compromise between localization in position and momentum is required to have states that evolve approximately classically for an extended period of time. But the uncertainty principle  $\Delta x \Delta p \geq \frac{\hbar}{2}$ famously limits the degree to which one can have both at the same time. For fixed $\hbar$, quantum states therefore must have some spread, which in turn implies that the approximate validity of classical equations for expectation values in generic background potentials cannot be made exact no matter what the state, and will always have some error associated with the finite widths of the state (except in the very special case of the harmonic oscillator, where expectation values evolve exactly classically for all times no matter what the state). While the absolute error in (\ref{EhrenfestApproxNR}) and (\ref{EhrenfestApproxSR}) can never be made to vanish for finite $\hbar$ by a particular choice of state because of the uncertainty principle, there is perhaps a sense in which the \textit{relative} error, such as the ratio of the absolute error to the magnitude of the classical terms in (\ref{EhrenfestApproxNR}) and (\ref{EhrenfestApproxSR}), can be made to vanish by focusing on progressively larger or more energetic systems. 

Feintzeig has recently proposed a ``factual" interpretation of the limit $\hbar \rightarrow 0$ in which changes to the numerical value of $\hbar$ are not interpreted counterfactually, but rather are induced by a change of units associated with descriptions of the system at increasingly large length scales. Feintzeig's focus is specifically on the relationship between classical and quantum algebras of observables reflected in the axioms of deformation quantization. By contrast, for the distinct type of formal correspondence considered here, in which quantum expectation values evolve approximately classically over a certain subset of the quantum state space, a mere change of units does not suffice in (\ref{EhrenfestApproxNR}) and (\ref{EhrenfestApproxSR}) to reduce the size of the ignored quantum correction terms relative to the classical terms that are kept, since both the neglected correction terms and classical terms have the same units, and therefore scale by the same factor under a change of units.

On the other hand, if we permit ourselves to vary $\hbar$ counterfactually, then in the limit $\hbar \rightarrow 0$ it is possible to have wave packets that are arbitrarily narrowly peaked in both position and momentum. In this case, there is no tradeoff between the quality of the approximation  in (\ref{EhrenfestApproxNR}) and the length of time for which these approximations hold. In the limit $\hbar \rightarrow 0$, it is possible for expectation values to satisfy classical equations of motion exactly by choosing states that are infinitely narrowly peaked in both position and momentum. However, the limit $\hbar \rightarrow 0$ \textit{by itself} does not imply that expectation values to satisfy classical equations of motion exactly, since it is also possible in this limit for the quantum state to have arbitrarily large widths in both position and momentum: in principle, nothing prevents $\Delta x \Delta p$ from being larger than $\frac{\hbar}{2}$, whatever the value of $\hbar$). In such a case,  expectation values will not evolve exactly classically in the limit $\hbar \rightarrow 0$, since there will still be some error arising from the finite widths of the wave packet (see \cite{Rosaler2018} for more detailed discussion of this point). In addition to the limit $\hbar \rightarrow 0$, one must also take the limit in which the position and momentum widths of the quantum state vanish as well. 

\subsection{Speculations on New Physics: Relating the $\textit{Standard Model} \rightarrow \textit{Quantum Gravity}$ and $\textit{General Relativity} \rightarrow \textit{Quantum Gravity}$ Reductions}

In the introduction, we speculated that a careful examination of the requirements for reduction involving more familiar theories might offer insight into the precise nature of the relationship that currently established theories 
\footnote{The terms ``theory" and ``model" are used loosely and interchangeably here.
}
such as the Standard Model (SM) and general relativity (GR) might bear to any theory of quantum gravity (QG) (where a theory of quantum gravity is understood here to encompass the domain of empirical validity not only of general relativity, but also of the Standard Model). 
\footnote{Of course, there also exist theories of quantum gravity, such as canonical approaches to quantum gravity, that make no claim to incorporate the successes of the Standard Model.
}
How, if at all, might the formal methodological considerations discussed above serve to clarify the nature of the relationship between current theories and any viable theory of quantum gravity? 

As Crowther has emphasized, any theory of quantum gravity, understood broadly as a quantum theory of spacetime, must \textit{by definition} reduce general relativity in the sense of recovering its empirical successes \cite{crowther2017inter}. Understood more narrowly as a quantum theory of spacetime that also recovers the successes of the Standard Model, any theory of QG must by definition reduce \textit{both} general relativity and the Standard Model. Whatever the correct unification of the Standard Model and general relativity turns out to be, it is widely expected to be a single mathematical structure - i.e., model - that encompasses all phenomena captured by these theories. Thus, although models in physics typically apply locally to specific systems, the universe itself is one example of such a system, and at least one vision of quantum gravity understands it to provide a single, cohesive model of this (extremely large) system, which contains all other subsystems. While some models apply only narrowly and locally, others, such as the Standard Model, are extremely broad in their scope; a model of quantum gravity would naturally lie at the very end of this spectrum in the latter direction.

\begin{figure} \label{QG}
\center{\includegraphics[width=.6 \linewidth]{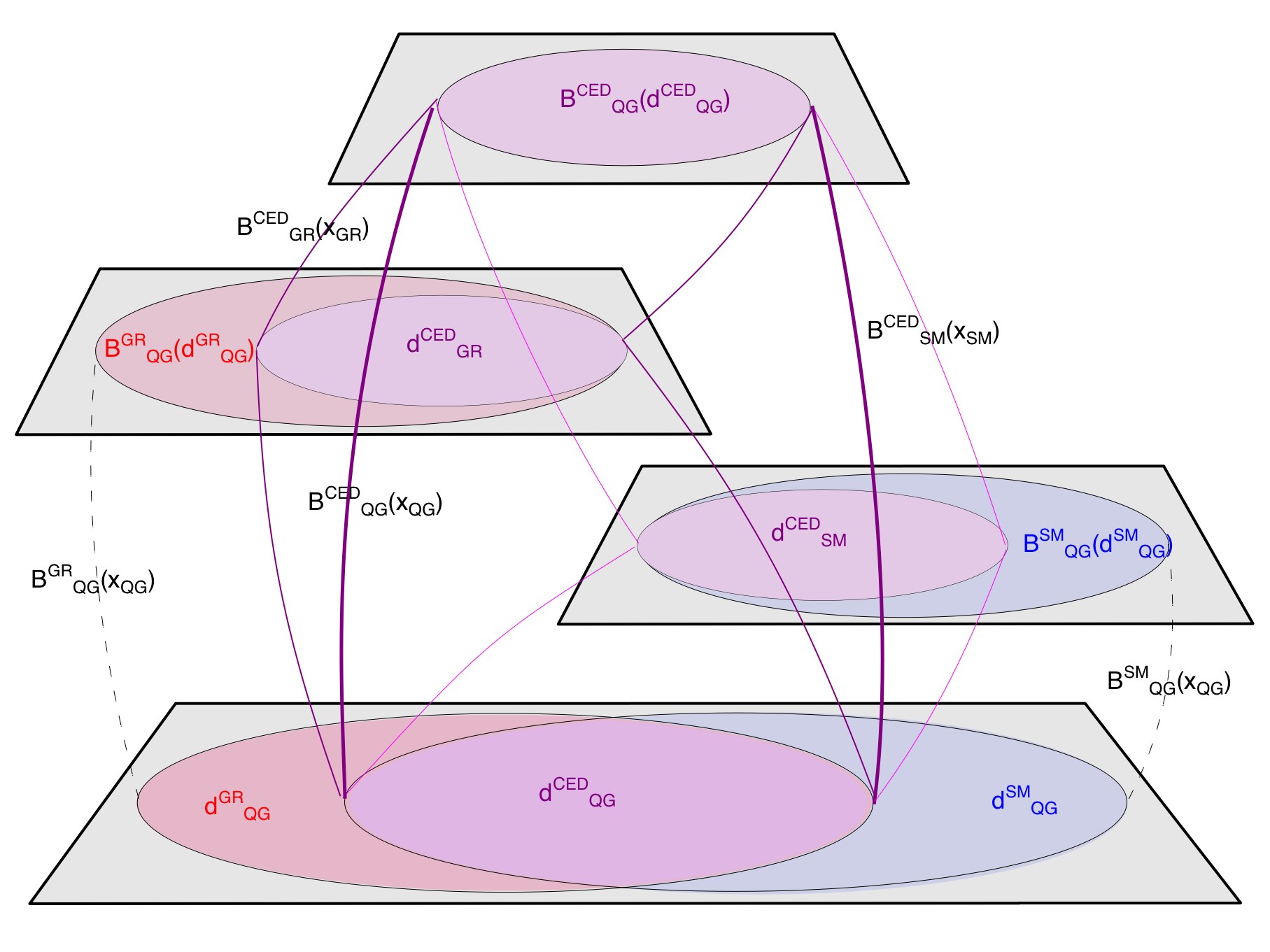}}
\caption{ Classical electrodynamics (CED) lies in the overlap of the domains of empirical validity of the Standard Model (SM) and general relativity (GR). It should therefore be possible to effect the reduction of classical electrodynamics (CED) to the correct model of quantum gravity (QG) - whatever that may be - either via the composite reduction $CED \rightarrow SM \rightarrow QG$ or via the composite reduction $CED \rightarrow GR \rightarrow QG$. The bridge function $B^{CED}_{QG}$ and state space domain $d^{CED}_{QG}$ should then be approximately independent of the path that one uses to perform the reduction. 
}
\label{BridgeFunctionQG}
\end{figure}

Applying the approach to reduction described above, it is reasonable to expect that there exist functions $B_{QG}^{SM}$ and $B_{QG}^{GR}$, and subsets $d_{QG}^{SM} \subset S_{QG}$ and $d_{QG}^{GR} \subset S_{QG}$, such that $B_{QG}^{SM}(x_{QG})$ approximately instantiates the equations of motion of $SM$
\footnote{This might be some quantum field theoretic Schrodinger equation for the Standard Model quantum state. 
}
for $x_{QG} \in d_{QG}^{SM}$ and such that $B_{QG}^{GR}(x_{QG})$ approximately instantiates the equations of motion of the $GR$ model for $x_{QG} \in d_{QG}^{GR}$. 
 In string theory, for example, $d^{GR}_{QG}$ is sometimes thought to include coherent states of the gravitational field, which are the quantum gravitational analogue to narrow wave packets considered in the context of ordinary quantum mechanics.  
\footnote{See, Huggett and Vistarini's \cite{huggett2014deriving}, and sources therein, for discussion of the role of coherent states in the relationship between string theory and classical general relativity.
}
Moreover, classical electrodynamics (CED) lies in the intersection of the domains of empirical validity of $SM$ and of $GR$ (presumed to include an electromagnetic term in its stress-energy tensor), so it should be possible as in the $NM \rightarrow RQM$ reduction to effect the reduction $CED \rightarrow QG$ via distinct composite reductions, $CED \rightarrow SM \rightarrow QG$ and  $CED \rightarrow GR \rightarrow QG$. 
\footnote{The relationship between GR and SM cannot be characterized as a case of reduction since it is not true that the domain of either is contained in that of the other. However, as we discuss here, their domains overlap. On this overlap domain, which includes the domain of CED, we expect a weaker relationship - what Crowther has called ``correspondence" - to hold. Correspondence requires distinct theories to approximately agree in cases where their domains of validity overlap. As Crowther emphasizes, reduction is a special case of correspondence in which the overlap of the theories' domains is the entire domain of one of the theories \cite{crowther2017inter}. 
}
Then, by analogy with the above discussion of the classical and non-relativistic domains of RQM, we should demand that 

\begin{align} \label{QGOverlap}
B^{CED}_{QG}(x_{QG}) \apeq B^{CED}_{SM}(B^{SM}_{QG}(x_{QG})) \approx B^{CED}_{GR}(B^{GR}_{QG}(x_{QG}))
\end{align}

\noindent for $x_{QG} \in d^{SM}_{QG} \cap d^{GR}_{QG} \subset S_{QG}$. This tells us that whatever $M_{QG}$ turns out to be, and whatever the state space domains of $d^{SM}_{QG} $ and $d^{GR}_{QG}$ turn out to be, $d^{SM}_{QG} $ and $d^{GR}_{QG}$ must overlap, and within this overlap the composite bridge functions  $B^{CED}_{GR}(B^{GR}_{QG}(x_{QG}))$ and $B^{CED}_{SM}(B^{SM}_{QG}(x_{QG}))$, constructed respectively via the compound reductions $CED \rightarrow SM \rightarrow QG$ and $CED \rightarrow GR \rightarrow QG$, should approximately satisfy the coupled Maxwell equations and Lorentz Force Law equations. This is illustrated pictorially in Figure \ref{BridgeFunctionQG}.






\section{Conclusion} \label{Conclusion}

Questions about the precise requirements for reduction between theories in physics bear on our understanding of the connections established theories and on efforts to construct new theories that recover the empirical successes of these established theories. I have critiqued one common approach, associated with the Bronstein cube of physical theories, as being either too vague or simplistic in its characterization of these requirements, and defended an alternative point of view based on the notion that reduction at its base concerns the relationship between distinct models of the same physical system, and that understanding the mechanisms of reduction between models requires identification of quantities in the reducing model that instantiate the physically salient structures of the reduced model. Using this simple strategy, it is possible to formalize the notion that reduction is transitive - i.e., that reductions can be chained together to directly relate the mathematical models used to describe increasingly remote realms of phenomena. Unlike presentations of the Bronstein cube approach to reduction, the approach described here articulates explicitly the sense in which quantum-to-classical and relativistic-to-non-relativistic transitions commute, as well as suggesting new ways in which current theories might serve to constrain the form of the mathematical laws governing new physics. For this reason, I hope that the reader will be convinced that careful attention to mathematical and methodological questions about the nature of reduction is worthwhile both from the perspective of understanding more deeply the connections among established theories and in the search for new, more encompassing theories. 

\

\noindent \textbf{Acknowledgments:} Thanks to Dennis Lehmkuhl for helpful discussions of reduction, and to Robert Harlander, Erhard Scholz, and the other members of the "Epistemology of the LHC" collaboration for feedback on earlier drafts of this work. Thanks also to audiences in Geneva, Salzburg, and Berlin. This research was supported by the DFG, grant FOR 2063. 
\def\bibsection{\section*{References}}
\bibliographystyle{plain}
\bibliography{References}

\end{document}